\documentclass[twocolumn,12pt,preprint]{aastex63}
\usepackage{amssymb}
\usepackage{epsfig}
\usepackage{natbib}
\usepackage{graphicx}
\usepackage{color}
\usepackage{tabularx}
\usepackage{epsfig}
\usepackage{rotating}
\usepackage{multirow}
\usepackage{subfigure}
\usepackage{ulem}
\usepackage{longtable}

\received{2019 November 29}
\revised{2020 May 18}
\accepted{2020 Jun 11}
\submitjournal{ApJS}

\shorttitle{The ZTF Catalog of Periodic Variable Stars}
\shortauthors{X. Chen et al.}

\begin{document}

\title{The Zwicky Transient Facility Catalog of Periodic Variable Stars}

\correspondingauthor{Xiaodian Chen}
\email{chenxiaodian@nao.cas.cn}

\author[0000-0001-7084-0484]{Xiaodian Chen}
\affiliation{CAS Key Laboratory of Optical Astronomy, National Astronomical Observatories, 
Chinese Academy of Sciences, Beijing 100101, China}
  
\author[0000-0003-4489-9794]{Shu Wang}
\affiliation{CAS Key Laboratory of Optical Astronomy, National Astronomical Observatories, 
Chinese Academy of Sciences, Beijing 100101, China}
   
\author[0000-0001-9073-9914]{Licai Deng}
\affiliation{CAS Key Laboratory of Optical Astronomy, National Astronomical Observatories, 
Chinese Academy of Sciences, Beijing 100101, China}

\author[0000-0002-7203-5996]{Richard de Grijs}
\affiliation{Department of Physics and Astronomy, Macquarie
  University, Balaclava Road, Sydney, NSW 2109, Australia}
\affiliation{Research Centre for Astronomy, Astrophysics and
  Astrophotonics, Macquarie University, Balaclava Road, Sydney, NSW
  2109, Australia}
\affiliation{International Space Science Institute--Beijing, 1
  Nanertiao, Zhongguancun, Hai Dian District, Beijing 100190, China}
    
\author[0000-0001-8247-4936]{Ming Yang}
\affiliation{IAASARS, National Observatory of Athens, Vas. Pavlou
  \& I. Metaxa, Penteli 15236, Greece}

\author[0000-0003-3347-7596]{Hao Tian}
\affiliation{Key Laboratory of Space Astronomy and Technology, National
  Astronomical Observatories, Chinese Academy of Sciences, Beijing 100101, China}

\begin{abstract}
The number of known periodic variables has grown rapidly in recent
years. Thanks to its large field of view and faint limiting magnitude,
the Zwicky Transient Facility (ZTF) offers a unique opportunity to
detect variable stars in the northern sky. Here, we exploit ZTF Data
Release 2 (DR2) to search for and classify variables down to
$r\sim20.6$ mag. We classify 781,602 periodic variables into 11 main
types using an improved classification method. Comparison with
previously published catalogs shows that 621,702 objects (79.5\%) are
newly discovered or newly classified, including $\sim$700 Cepheids,
$\sim$5000 RR Lyrae stars, $\sim$15,000 $\delta$ Scuti variables,
$\sim$350,000 eclipsing binaries, $\sim$100,000 long-period variables,
and about 150,000 rotational variables. The typical misclassification
rate and period accuracy are on the order of 2\% and 99\%,
respectively. 74\% of our variables are located at Galactic latitudes,
$|b|<10^\circ$. This large sample of Cepheids, RR Lyrae, $\delta$
Scuti stars, and contact (EW-type) eclipsing binaries is helpful to
investigate the Galaxy's disk structure and evolution with an improved
completeness, areal coverage, and age resolution. Specifically, the
northern warp and the disk's edge at distances of 15--20 kpc are
significantly better covered than previously. Among rotational
variables, RS Canum Venaticorum and BY Draconis-type variables can be
separated easily. Our knowledge of stellar chromospheric activity
would benefit greatly from a statistical analysis of these types of
variables.
\end{abstract}

\section{Introduction}

Periodic or quasi-periodic variables encompass mainly eclipsing binary
systems and rotational stars characterized by extrinsic variability,
while pulsating variables are among objects exhibiting intrinsic
fluctuations in the variability tree
\citep{2019A&A...623A.110G}. These objects are interesting, since
their periods contain information about their luminosity or
mass. Variable stars have been studied for several hundred
years. However, this field has been advancing significantly in recent
years. With increases in the depth and efficacy of time-domain
surveys, the number of newly detected variables is roughly doubled
every year. {\sl Gaia}'s third Data Release (DR3) is expected to
include 7 million variables, with about one-third to half of periodic
nature. This huge number is of the same order as the number of extant
stellar spectra. By combining the spectra and light curves (LCs) of
variable stars, we will be able to improve our knowledge of a range of
fundamental stellar parameters, which in turn can benefit developments
in theory and the applications of stellar physics. A complete catalog
of all-sky periodic variables can also be used to separate variable
and non-variable stars in spectroscopic surveys---such as those
undertaken with the Large Sky Area Multi-Object Fiber Spectroscopic
Telescope \citep[LAMOST;][]{2012RAA....12.1197C, 2012RAA....12..735D}
and the Sloan Digital Sky Survey
\citep[SDSS;][]{2011AJ....142...72E}---to reduce the systematic
effects pertaining to the presence of variables.

Although searches for variables date back hundreds of years, the
number of known variables first increased significantly in the 1990s
thanks to the development of CCDs. The Massive Compact Halo Object
(MACHO) survey \citep{1993Natur.365..621A} found 20,000 variable stars
in the Large Magellanic Cloud (LMC). The Optical Gravitational Lensing
Experiment (OGLE) subsequently made significant progress in this
field. Over its 20-year duration, OGLE detected more than 900,000
variables in the Magellanic Clouds, the Galactic bulge, and the
Galactic plane \citep{1992AcA....42..253U, 2015AcA....65....1U}. The
first all-sky variability survey was carried out by the All Sky
Automated Survey (ASAS), which detected 10,000 eclipsing binaries and
8000 periodic pulsating stars \citep{2005AcA....55..275P}. Several
thousand variables were also found by the Robotic Optical Transient
Search Experiment \citep[ROTSE;][]{2000AJ....119.1901A} and the
Lincoln Near-Earth Asteroid Research
\citep[LINEAR]{2013AJ....146..101P} project. The Catalina Surveys
\citep{2014ApJS..213....9D, 2017MNRAS.469.3688D} found more than
100,000 variables, covering almost the entire sky, which allowed
significant progress to be made in studying streams, structures, and
the shape of the Milky Way's halo. Variable surveys have also been
undertaken in infrared passbands, e.g., the VISTA Variables in the
V\'{i}a L\'{a}ctea (VVV) survey \citep{2010NewA...15..433M} and the
VISTA survey of the Magellanic Clouds system
\citep[VMC;][]{2011A&A...527A.116C}. In 2018, a number of
variable-star catalogs were published, including the All-Sky Automated
Survey for Supernovae \citep[ASAS-SN;][]{2018MNRAS.477.3145J}, the
Asteroid Terrestrial-impact Last Alert System
\citep[ATLAS;][]{2018AJ....156..241H}, the Wide-field Infrared Survey
Explorer ({\sl WISE}) catalog of periodic variable stars
\citep{2018ApJS..237...28C}, and the variable catalog of {\sl Gaia}
DR2 \citep{2019A&A...622A..60C, 2018A&A...618A..58M}. The number of
known variables is currently experiencing a second substantial
increase, resulting in millions of known variable objects.

Alongside the increasing numbers of known variables, advances have
also been made as regards the corresponding and ancillary science,
e.g., pertaining to distance measurements, structure analysis,
stellar-mass black hole searches, and multi-messenger astrophysics. As
regards distance indicators, detached eclipsing binaries were used to
measure the distance to the LMC to an accuracy of 1\%
\citep{2019Natur.567..200P}. This accurate benchmark distance is of
great importance to refine the value of the Hubble parameter
\citep{2019ApJ...876...85R}. As to structure analyses, the sample of
newly found classical Cepheids led to the first intuitive 3D map of
two-thirds of the Milky Way's disk \citep{2019NatAs...3..320C,
  2019Sci...365..478S}. This new map will help us understand the
evolution of both the Galaxy's spiral arms and its outer disk. As
regards long-period binary variables, their unseen companions may
include black holes. A low-mass black hole has been found to orbit a
giant star with an orbital period of $\sim$83 days
\citep{Thompson637}. Searching for variables with the shortest or the
longest periods is attractive \citep{2019Natur.571..528B,
  2015Natur.518...74G}, since such systems contain information about
both the properties of their electromagnetic radiation and possible
gravitational-wave signals that can motivate multi-messenger
astrophysics.

In this paper, we search for periodic variables in the recent data
release of the Zwicky Transient Facility (ZTF), the deepest
time-domain survey of the northern sky. Periodicities are analyzed for
the full database to identify a sample of periodic variables that is
as complete as possible. Using an improved method of classification,
781,602 variables belonging to 11 main types are classified. Our
sample's detection rate of newly identified objects is about 80\%. We
construct a large variable-star catalog that will benefit many
follow-up studies. The data are described in Section 2. We explain how
we identify and classify variables in Sections 3 and 4,
respectively. The new variable catalog and a comparison of its
consistency with previous catalogs are covered in Section 5. We
provide a detailed discussion of each type of variable object and
their applications in Section 6. Section 7 summarizes our conclusions.

\section{ZTF data}

The ZTF is a 48-inch Schmidt telescope with a 47 deg$^2$ field of view
\citep{2019PASP..131a8003M, 2019PASP..131f8003B}. This large field of
view ensures that the ZTF can scan the entire northern sky every
night. The ZTF survey started on 2018 March 17. During the planned
three-year survey, ZTF is expected to acquire $\sim$450 observational
epochs for 1.8 billion objects. Its main science aims are the physics
of transient objects, stellar variability, and solar system science
\citep{2019PASP..131g8001G, 2019PASP..131c8002M}. ZTF DR2 contains
data acquired between 2018 March and 2019 June, covering a timespan of
around 470 days. ZTF DR2 photometry is obtained in two observation
modes, three-night cadence for the northern sky and one-night cadence
for the Galactic plane, $|b| \le 7^\circ, {\rm Dec.}>-25^\circ$. The
photometry is provided in the $g$ and $r$ bands, with a uniform
exposure time of 30 s per observation. The limiting magnitude is $r
\sim 20.6$ mag. ZTF $g$ and $r$ photometry is calibrated with the help
of Pan-STARRS1 DR1 \citep{2016arXiv161205560C}; there is only a small
offset of $\sim 0.01$ mag between both systems at the bright end
($r<15.5$ mag). ZTF DR2 includes more than 1 billion stars, about half
of which have $>20$ epochs of observations. For the majority of stars
located in the northern Galactic plane, ZTF contains $\sim150$ epochs
of observations. As such, ZTF can be used to detect numerous variables
in the northern Galactic plane, which has not been well-studied by
previous time-domain surveys. We downloaded the full catalog
(comprising a total volume of $\sim$3.4
TB),\footnote{\url{https://irsa.ipac.caltech.edu/data/ZTF/lc_dr2/}}
which is provided in the form of 856 files, ordered as a function of
spatial position.

\section{Search for periodic variables}

To search for periodic variables, we need to impose some conditions to
make the process easier and quicker. First, we only selected objects
in the ZTF data with at least 20 detections. Adoption of this
criterion will not cause us to miss many periodic candidates, since a
period's false-alarm probability (FAP) based on $<20$ detections is
high. Second, poor-quality images and photometry were excluded by
adopting ${\rm INFOBITS}<33,554,432$ and ${\rm catflags}\neq 32,768$,
respectively. Third, preliminary searching for periodicity for a given
object was based on photometry associated with the same internal
product ID. This choice made this process much quicker. Although a
fraction of the objects have multiple internal IDs, since they may
have been observed for different projects, their periodicity is
unlikely to be missed based on analysis of the data from the main
project. A redetermination of the period will be done based on the
object's position (R.A. and Dec.) during the next steps.

We cut the data into smaller segments to mitigate computational memory
restrictions and ran Lomb--Scargle periodogram
\citep{1976Ap&SS..39..447L, 1982ApJ...263..835S} analysis using the
MATLAB code `plomb' to search for variables. The code returns the
Lomb--Scargle power spectral density (PSD) based on the maximum input
frequency. We searched for periods from 0.025 to 1000 days (in steps
of 0.0001 day$^{-1}$ in frequency), a range covering variables from
short-period $\delta$ Scuti to long-period Mira stars. We attempted to
exclude aliased periods---such as one day and its multiples---using a
method similar to that of \citet{2019MNRAS.486.1907J}. Some aliased
periods were successfully excluded, although many remained. Many
candidate variables with aliased periods are real long-period
variables (LPVs). Their aliased periods are a combination of real
periods, $P_{\rm r}$, and the one-day sampling cadence, $P_{\rm c}$:
$1/P_{\rm a}=1/P_{\rm r}+1/P_{\rm c},P_{\rm c}=0.5, 1, 2 ...$. After
excluding $P_{\rm a}$, the real periods of well-sampled variables
could be obtained (see also Section 5.3).

A number of parameters were recorded during this process to help with
the selection of variable candidates, including the FAP, which
represents the confidence of the periodicity determination. We
selected periodic candidates by imposing ${\rm FAP} < 0.001$. Adoption
of this criterion mistakenly excludes only a few short-period
variables. These will be rediscovered with better sampling in the
ZTF's future DRs. Other parameters of importance are the mean PSD,
$\langle P_f \rangle$, and the number of frequencies associated with
${\rm PSD}>0.5 {\rm PSD}_{\rm max}$. These parameters are used to
avoid inclusion of false variables produced by abnormal sampling. ZTF
is not affected by problems caused by the inhomogeneous distribution
of the data points, except for LPVs. Therefore, we do not need to
record any parameters to trace the distribution of the data points in
the phase-folded LC. Periodicities were determined in both the $g$ and
$r$ bands, and candidates were accepted if the LC in one band met all
conditions.

Running the code took one month on two computers with a total of 18 i7
CPUs. A grand total of 1.4 million variable candidates were
recorded. We redetermined their periods based on all of the
photometric data within $1''$ around their spatial position. The LCs
were fitted with a fourth-order Fourier function, $\displaystyle{f
  =a_0 + \sum_{i=1}^4a_i\cos(2\pi it/P+\phi_i)}$, which is a suitable
choice for survey LCs to avoid missing important parameters or
overfitting. About 90\% of these ZTF LCs have $a_4$ values less than
the typical photometric uncertainties of 0.02 mag. $a_i$ and $\phi_i$
denote amplitudes and phases for each order, respectively. Candidates
with poorly fitted LCs were excluded using the adjusted $R^2$ (which
represents how well LCs are fitted by the Fourier function), i.e.,
$R_g^2<0.4$ and $R_r^2<0.4$. Objects still characterized by aliased
periods were excluded based on their distributions in the period
versus $R^2$ or $\phi_{21}$ ($\phi_{21}=\phi_{2}-2\phi_{1}$) diagram,
e.g., objects with $|P-1|<0.03$ days and $R^2<0.8$ or
$|{\phi_{21}}_r-{\phi_{21}}_g|>0.2$ were excluded. At this point, 1
million candidates remained for further classification.

\section{Classification}

Classification is more challenging than the identification
process. The main idea is based on the varying densities of different
variables in multi-dimensional space. Parameters including the period
($\log P$), phase difference ($\phi_{21}$), amplitude ratio ($R_{21}
=a_2/a_1$), amplitude (${\rm Amp.}$), absolute Wesenheit magnitude
($M_{W_{gr}}$), and adjusted $R^2$ were adopted to help with our
classification. The period is the most significant parameter;
$\phi_{21}$, $R_{21}$, and ${\rm Amp.}$ are parameters determined from
the LC fits. The absolute Wesenheit magnitude is used to separate the
variables in the overlap region of the LCs' parameter space. It is
also helpful to narrow down the types of low-amplitude variables. To
avoid significant extinction in the optical bands, we calculated
Wesenheit magnitudes, $m_{W_{gr}}=\langle g \rangle
-R_{gr}\times(\langle g \rangle - \langle r \rangle)$, making use of
the $g,r$ photometry. Here, $\langle g \rangle$ and $\langle r
\rangle$ are the mean magnitudes determined from our LC fits. We
adopted the recently derived Galactic extinction law
\citep{2019ApJ...877..116W} based on red clump stars with accurate
parameters and obtained $m_{W_{gr}}=3.712 \langle r \rangle -2.712
\langle g \rangle$.

{\sl Gaia} DR2 parallaxes \citep{2018AA...616A...1G} were adopted to
estimate the absolute magnitudes. Since many of our objects are
located in the Galactic plane, we used the inverse of the parallaxes
to estimate distances. To reduce the systematic effects of the
Lutz--Kelker bias \citep{1973PASP...85..573L}, only parallaxes with
$\varpi>0.2, \sigma_{\varpi}/\varpi<0.2$ and
$\sigma_{\varpi}/\varpi<0.5$ were adopted for our detailed
classification and the dwarf--giant separation, respectively. $R^2$
contains information to separate characteristic and non-characteristic
LCs, where `characteristic' LCs have similar patterns, and we can find
a rule to classify variable stars based on their LCs.
`Non-characteristic' LCs exhibit arbitrary patterns. In other words,
variables with characteristic LCs form either a clump or a sequence in
the $P$ versus $\phi_{21}$ diagram, while variables with
non-characteristic LC exhibit a random distribution. As $R^2$
decreases, the probability of variables with characteristic
LCs---e.g., eclipsing binaries and radially pulsating
stars---decreases.

Note that the relation between the amplitudes in both filters (${\rm
  Amp}_g=a\times {\rm Amp}_r+b$) is different for each variable. The
1$\sigma$ scatter (0.03--0.06 mag) on these relations in the current
database is too large to allow for a robust classification of the
majority of our variables, considering that the amplitude difference
between the $g$ and $r$ bands is not significant. As such, we did not
adopt this criterion for automatic classification, but only for a
visual check of the easily-confused variables.

The classification process proceeded based on the auto-manual
Density-Based Spatial Clustering of Applications with Noise
(DBSCAN). DBSCAN clusters data points based on a neighborhood search
radius and a minimum number of neighbors. Both of these parameters
were adjusted in each step to achieve an optimal separation of the
different variables. Since classification rules for variables (in
particular for variables with non-characteristic LCs) are not
well-established, we only classified variables with a significant
density in parameter space. The classification order progresses from
Cepheids, to fundamental-mode (RRab) and first-overtone (RRc) Lyrae,
$\delta$ Scuti stars, eclipsing binaries, and finally to variables
with non-characteristic LCs.

As our first step, we established the best samples of Cepheids, RRab
and RRc Lyrae, and $\delta$ Scuti stars based on strict constraints on
the LC quality ($R^2>0.9$, similar periods in both filters), the LC
shape, and stellar luminosity. For each type of variable, we found
around 1000 candidates. These objects were added to the remaining
candidates to enhance their density in parameter space. We then
selected Cepheids, RRab and RRc Lyrae, and $\delta$ Scuti stars based
on their density of data points. In doing so, candidates located close
to the best-established sample objects were assumed to be of the same
type. This process was repeated until the density of the remaining
candidates was similar to the density of variables with
non-characteristic LCs in their vicinity. After removing these four
types of variables, the majority of the remaining candidates were
eclipsing binaries. We selected all eclipsing binaries with amplitudes
larger than 0.08 mag to distinguish them from RS Canum
Venaticorum-type systems (RS CVn), which are eclipsing binaries
exhibiting chromospheric activity.

The LCs of eclipsing binaries are highly symmetric, and their phase
differences do not deviate significantly from $2\pi$
($|\phi_{21}-2\pi|<0.1\pi$) for periods longer than 0.3 days (half
periods for eclipsing binaries\footnote{The period determination
  process produces half-periods for the majority of eclipsing binaries
  because the two eclipses over a full period are too similar to be
  separated. Only very few eclipsing binaries with significantly
  modulated LCs yield full periods.}). However, contact (EW-type)
eclipsing binary systems with half-periods shorter than 0.3 days show
a more scattered distribution of $\phi_{21}$. We retained these
half-periods until we had completed the separation of eclipsing
binaries from other types of variables. Since the number density of
EW-type eclipsing binaries is higher than that of the other types of
periodic variables \citep[around 0.1--0.4\% of the overall stellar
  population, depending on environment][]{2006MNRAS.368.1319R,
  2016AJ....152..129C}, more than half of our candidates are eclipsing
binaries. After removal of the eclipsing binaries, the remaining
candidates were variables characterized by non-characteristic
LCs. From short to long periods, they are low-amplitude $\delta$ Scuti
(LAD) stars, RS CVn, BY Draconis (BY Dra)-type variables, semi-regular
variables (SRs), and Miras. LADs have shorter periods and brighter
luminosities than their counterpart EW-type eclipsing binaries or RS
CVn.

The LCs of RS CVn show an additional sine signal out-of-eclipse, which
is the result of chromospheric activity. They have very similar
luminosities to eclipsing binaries. BY Dras are K--M dwarf rotational
stars also exhibit chromospheric activity. Their rotational periods
range from less than one to a few days. Since almost all K--M dwarfs
show BY Dra LCs, they are second in terms of their numbers only to
EW-type eclipsing binaries in magnitude-limited samples. The period
ranges of BY Dra and RS CVn overlap significantly; they can only be
separated with the help of their luminosities. BY Dra are fainter than
RS CVn. Their absolute magnitudes are concentrated in the range
$2.8<M_{W_{gr}}<4.6$ mag. We also found that the luminosities of BY
Dra have no relation to their periods, which means that stellar mass
does not affect the rate of rotation among these objects. SRs are red
giants, red supergiants, or asymptotic giant branch (AGB) stars with
periodic and variable LCs. SR periods range from 20 to thousands of
days, which overlaps significantly with the Mira period range. Miras
are oxygen- or carbon-rich AGB stars. The DBSCAN results show that SRs
and Miras can be separated because of an apparent difference in
amplitude \citep[see also][]{2018MNRAS.477.3145J}. Here, we selected
Miras by adopting $\rm {Amp}_r>2$ mag or $\rm {Amp}_g>2.4$.

\begin{figure}[h!]
\centering
\vspace{-0.0in}
\includegraphics[angle=0,width=85mm]{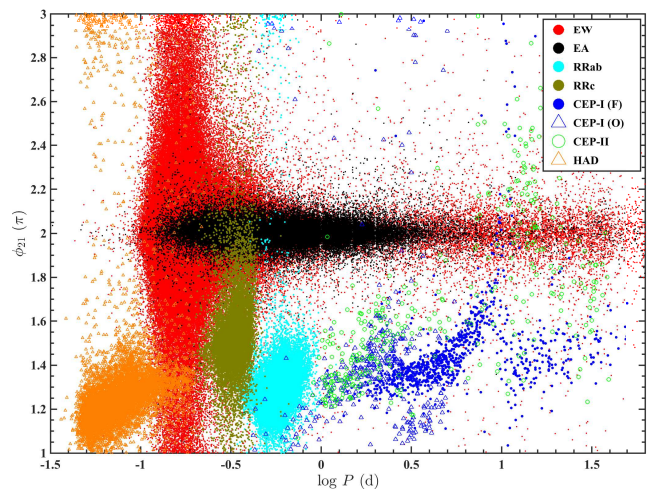}
\vspace{-0.0in}
\caption{\label{cla1}$\phi_{21}$ versus $\log P$ diagram for all ZTF
  variables with characteristic LCs. From short to long periods, these
  include high-amplitude $\delta$ Scuti stars (HADs: orange), EW-type
  eclipsing binaries (red), RRc Lyrae (dark green), RRab Lyrae (cyan),
  EA-type eclipsing binaries (black), first-overtone classical
  Cepheids (blue triangles), fundamental-mode classical Cepheids (blue
  dots), and Type II Cepheids (green circles).}
\end{figure}

\begin{figure}[h!]
\centering
\vspace{-0.0in}
\includegraphics[angle=0,width=85mm]{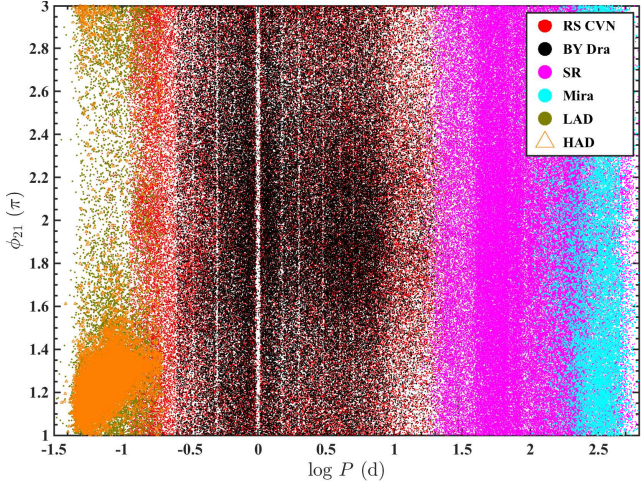}
\vspace{-0.0in}
\caption{\label{cla2}$\phi_{21}$ versus $\log P$ diagram for all ZTF
  variables with non-characteristic LCs. From short to long periods,
  these include low-amplitude $\delta$ Scuti stars (LADs: dark green),
  RS CVn (red), BY Dra stars (black), SRs (magenta), and Miras
  (cyan). Orange HADs are also shown, for comparison with the LADs.}
\end{figure}

The criteria adopted to assist with the DBSCAN process are listed in
Table \ref{T1}. The $\phi_{21}$--$\log P$ diagrams of all variables
are shown in Figures \ref{cla1} and \ref{cla2}. We find that aliased
periods around one day are almost excluded. In Figure \ref{cla2},
variables with non-characteristic LCs exhibit a random distribution of
$\phi_{21}$ between $\pi$ and $3\pi$. A low density in $-0.7<\log
P<1.3$ [days] and $1.95\pi<\phi_{21}<2.05\pi$ is due to the difficulty
to separate non-characteristic LCs from eclipsing binaries' LCs.

Eclipsing binaries can be divided into EW, semi-detached (EB), and
detached (EA) types according to the degree by which they fill their
Roche lobes. An accurate classification requires a detailed orbital
analysis, which represents a major effort. Fortunately, we can also
use well-established empirical relations related to their LCs. We
refitted the LCs of our sample of eclipsing binaries using their full
periods. The newly determined Fourier parameters $a_2$ and $a_4$
(similar to $a_1$ and $a_2$ pertaining to their half-periods) and the
difference between the primary and secondary minimum magnitudes
$\Delta_{\rm{min}}=m_{\rm p}-m_{\rm s}$ were obtained to perform the
classification. Compared with EW-type eclipsing binaries, EA-type
objects have a larger amplitude $a_4$. To separate EW from EA types,
we used the ($a_4, a_2$) diagram, similarly to
\cite{2018ApJS..237...28C}.

We also adopted $R_{41}=a_4/a_2=0.43$ (see Figure \ref{pr21}) as
additional boundary to identify eclipsing binaries that were not well
separated based on the equations of \cite{2018ApJS..237...28C}. The
main difference between semi-detached and contact eclipsing binaries
is the difference between both minima. Both components of contact
eclipsing binary systems share a common envelope and, as a result,
they have similar temperatures. However, semi-detached eclipsing
binaries usually exhibit larger $\Delta_{\rm{min}}$ because of the
different temperatures of the two components. $\Delta_{\rm{min}}=0.2$
mag (corresponding to a 5\% difference between the components'
temperatures for an inclination of $i=90^\circ$) is usually adopted as
the boundary between semi-detached and contact eclipsing
binaries. This criterion offers a rough classification of the two
subtypes, and it is reliable for higher orbital inclinations
($i>70^\circ$) and good photometric quality. We did not try to
classify and distinguish semi-detached eclipsing binaries from other
types of eclipsing binaries. Instead, we recorded the two-band minimum
differences $\Delta_{\rm{min}}$ in our catalog to facilitate any
further classification based on future DRs.

\begin{figure}[h!]
\centering
\vspace{-0.0in}
\includegraphics[angle=0,width=85mm]{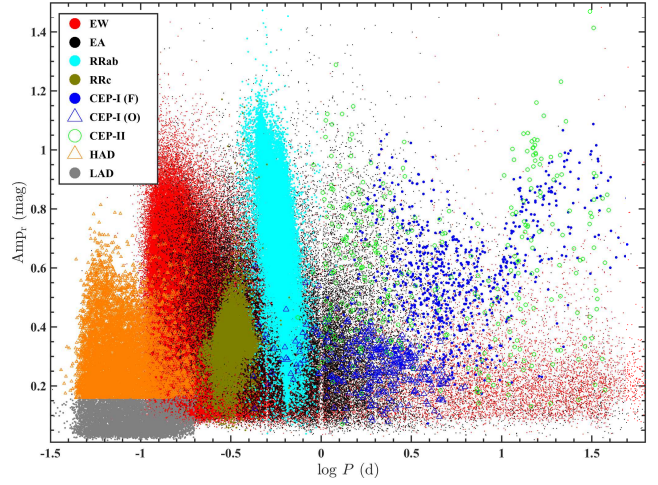}
\vspace{-0.0in}
\caption{\label{pa}${\rm{Amp}}_r$ versus $\log P$ diagram for all ZTF
  variables. Symbols are as in Figure \ref{cla1}. LADs have been added
  as gray dots.}
\end{figure}

\begin{figure}[h!]
\centering
\vspace{-0.0in}
\includegraphics[angle=0,width=85mm]{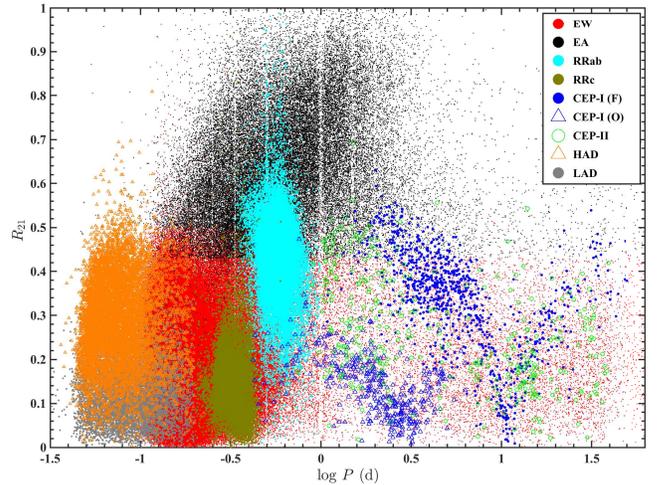}
\vspace{-0.0in}
\caption{\label{pr21}$R_{21}$ versus $\log P$ diagram for all ZTF
  variables. Symbols are as in Figure \ref{pa}.}
\end{figure}

The main subtypes of Cepheids among our candidates are classical
Cepheids pulsating in their fundamental and first-overtone modes and
Type II Cepheids. Classical and Type II Cepheids cannot be separated
cleanly only based on LC information. However, classical Cepheids are
two or three magnitudes brighter than Type II Cepheids. In addition,
classical Cepheids are young and associated with the Milky Way's thin
disk. Armed with these two additional classification criteria, both
Cepheid types can be separated more easily. The fundamental and
first-overtone classical Cepheids exhibit different LC
shapes. First-overtone Cepheids usually have smaller amplitudes (${\rm
  Amp.}~<0.4$ mag) and a low amplitude ratio ($R_{21}<0.3$). In
Figures \ref{pa} and \ref{pr21}, we can see a clear boundary between
fundamental (blue dots) and first-overtone (blue triangles) classical
Cepheids \citep[see also][]{2008AcA....58..163S}. Just like for
Cepheids, RRab and RRc Lyrae pulsating in their fundamental and
first-overtone modes also exhibit a clear boundary.

\subsection{Luminosity Selection}

\begin{figure}[ht!]
\centering
\vspace{-0.0in}
\includegraphics[angle=0,width=85mm]{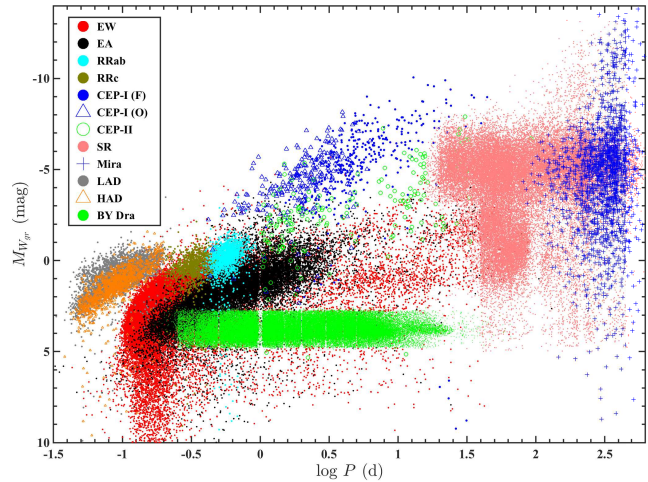}
\vspace{-0.0in}
\caption{\label{pl} $\log P$ versus $M_{W_{gr}}$ diagram for variables
  with better than 20\% (all variables except giants), 50\% (Miras and
  SRs), and 100\% (Cepheids) {\sl Gaia} parallax
  accuracy. $M_{W_{gr}}$ are the absolute Weisenheit magnitudes
  estimated from a combination of $g, r$ photometry and {\sl Gaia}
  parallaxes. Symbols are as in Figure \ref{pr21}. Miras, SRs, and BY
  Dra variables have been added using blue plus signs, and pink and
  green dots, respectively.}
\end{figure}

Luminosity is the other important parameter one can use to classify
variables. We present the period--luminosity diagram in Figure
\ref{pl}. Variables with {\sl Gaia} parallax accuracy better than 20\%
are shown, except for the luminous variables, since most of the latter
are distant and exceed our parallax constraint. Cepheids with $<100$\%
parallax uncertainty, Miras with $<50$\% parallax uncertainty, and SRs
with $<50$\% parallax uncertainty are shown. Variables outside these
ranges exhibit more scatter in their absolute magnitude distributions
and a tail at faint magnitudes, which is the result of the
Lutz--Kelker effect.

$\delta$ Scuti stars, RRab and RRc Lyrae, and classical Cepheids that
are located in the instability strip follow tight PLRs. The scatter in
their absolute magnitudes is mainly caused by parallax
uncertainties. We adopted $M_{W_{gr}}-M_{W_{gr}}({\rm
  PL})<3\sigma_{\rm DM}$ to exclude dwarf contamination. Here,
$M_{W_{gr}}$ and $M_{W_{gr}}({\rm PL})$ are the observed and predicted
absolute magnitudes, respectively, and $\sigma_{\rm DM}$ is the error
in the distance modulus, which has been propagated from the parallax
uncertainty. The $M_{W_{gr}}({\rm PL})$--$P$ relations for these
variables were calculated using objects with parallax uncertainties
less than 10\%. Since these PLRs were roughly determined and only used
to exclude contamination, we do not report the corresponding
coefficients here. For more accurate PLRs, we refer to individual
papers discussed in Section 6.

In Figure \ref{pl}, we find that both HADs and LADs follow similar
PLRs, and their scatter seems smaller than that pertaining to the RR
Lyrae and Cepheids. This smaller scatter is driven by the high number
of nearby $\delta$ Scuti stars. Except for these variables, EW-type
eclipsing binaries, Type II Cepheids, SRs, and Miras also follow PLRs
characterized by moderate scatter. EW-type eclipsing binaries have
been found to obey PLRs with 6--7\% accuracy in infrared bands
\citep{2016ApJ...832..138C, 2018ApJ...859..140C}. SRs and Miras are
LPVs that obey different sequences in the period--luminosity diagram
\citep{2019A&A...631A..24L}. Without access to accurate parallaxes and
well-determined periods, their sequences are rather unclear.

At the bottom of the the period--luminosity diagram, we find that BY
Dra variables are vertically distributed in the range of
$2.8<M_{W_{gr}}<4.6$ mag, and they are fainter than other
variables. This BY Dra sequence is always clear in a parallax-limited
sample, which means that this is a real distribution rather than one
caused by selection effects associated with the limit imposed on the
{\sl Gaia} parallaxes. Since the absolute magnitude boundary for BY
Dra variables is unknown, we adopted these rough luminosity cuts to
classify BY Dra stars. The rotational periods of BY Dra variables
range from 0.25 to 20 days, with a peak around a few days. RS CVn
occupy the same distribution as eclipsing binaries. They are not shown
in the diagram. The PLRs or luminosity ranges were determined for each
variable type to help with our classification. For example, candidates
2 mag fainter than classical Cepheids have a low probability of being
classical Cepheids..

For candidates in the overlap region in parameter space, such as RRc
and EW-type eclipsing binaries, Cepheids and rotational variables,
LADs and EW-type eclipsing binaries, we performed a visual check of
their LCs to assist with and confirm the classification process. We
thus classified 781,602 periodic variables from our master sample of
2.1 million candidates. The remaining candidates were recorded as
suspected variables. They will be better classified based on enhanced
data from future ZTF DRs.

\begin{table*}[ht!]
\begin{center}
\caption{\label{T1}Criteria adopted to classify different types of variables}
\begin{tabular}{ll}
\hline
\hline
                             Type& Selection criteria \\                                                                 
\hline                                                                                                                   
LAD              & $R^2>0.4$, ${\rm Amp.}<0.2$,  ${\rm FAP}<0.001$,  $M_{W_{gr}}-M_{W_{gr}}({\rm PL})<3\sigma_{\rm DM}$, $P<0.2$ days        \\                 
HAD              & $R^2>0.4$, ${\rm Amp.}>0.2$,  ${\rm FAP}<0.001$,  $M_{W_{gr}}-M_{W_{gr}}({\rm PL})<3\sigma_{\rm DM}$           \\             
RRc              & $R^2>0.6$, ${\rm Amp.}>0.08$, ${\rm FAP}<0.001$,  $M_{W_{gr}}-M_{W_{gr}}({\rm PL})<3\sigma_{\rm DM}$, $\phi_{21}<1.9$\\                
RRab             & $R^2>0.6$, ${\rm Amp.}>0.08$, ${\rm FAP}<0.001$,  $M_{W_{gr}}-M_{W_{gr}}({\rm PL})<3\sigma_{\rm DM}$ \\                            
Cepheid          & $R^2>0.6$, ${\rm Amp.}>0.08$, ${\rm FAP}<0.001$,  $M_{W_{gr}}-M_{W_{gr}}({\rm PL})<3\sigma_{\rm DM}$, $P<40$ days \\                              
Eclipsing binary & $R^2>0.4$, ${\rm Amp.}>0.08$, ${\rm FAP}<0.001$                                \\                                                              
Mira             & $R^2>0.4$, ${\rm Amp.}>=2$,   ${\rm FAP}<0.001$ , $P>80$ days                                 \\                      
SR               & $R^2>0.4$, ${\rm Amp.}<2$,    ${\rm FAP}<0.001$, $P>20$ days                                \\
BY Dra           & $R^2>0.4$, ${\rm FAP}<0.001$, $2.8<M_{W_{gr}}<4.6$,  $0.25<P<20$ days                      \\
RS CVn           & $R^2>0.4$, ${\rm FAP}<0.001$, $M_{W_{gr}}~=M_{W_{gr}}$(Eclipsing  binary), $P<20$ days     \\      
\hline                                                                                                                   
EW               & $a_4<=a_2(a_2+0.125)$                                               \\
                 & $a_2(a_2+0.125)<a_4<a_2(a_2+0.375)$,  $a_4/a_2<=0.43$        \\
EA               & $a_4>=a_2(a_2+0.375)$                                                      \\                          
                 & $a_2(a_2+0.125)<a_4<a_2(a_2+0.375)$,  $a_4/a_2>0.43$                           \\                          
                                                                                                                         
\hline                                                                                                                   
\end{tabular}                                                                                                            
\end{center}                                                                                                             
\end{table*}    

\begin{table*}[ht!]
\tiny
\vspace{-0.0in}
\begin{center}
\caption{\label{t2}ZTF Variables Catalog\tablenotemark{a}.}
\vspace{0.15in}
\begin{tabular}{lcccccccccccl}
\hline
\hline
ID  & R.A. (J2000)    & Dec. (J2000)   & Period  & $R_{21}$ & $\phi_{21}$ &$T_0$ &$\langle g \rangle$ & $\langle r \rangle$    &...&Amp$_g$&Amp$_r$& Type  \\      
    & $^\circ$ & $^\circ$ & days    &          &              & HJD$-$2400000.5     & mag  & mag      &...& mag        & mag        &        \\
\hline     
ZTFJ000000.13+620605.8 &  0.00056 & 62.10163 & 1.9449979  & 0.198& 5.189& 58386.2629478& 17.995& 16.571&...  & 0.113& 0.078&  BYDra\\
ZTFJ000000.14+721413.7 &  0.00061 & 72.23716 & 0.2991500  & 0.263& 6.308& 58388.2555794& 19.613& 18.804&...  & 0.540& 0.438&  EW   \\
ZTFJ000000.19+320847.2 &  0.00080 & 32.14645 & 0.2870590  & 0.010& 8.024& 58280.4780813& 15.311& 14.610&...  & 0.219& 0.197&  EW   \\
ZTFJ000000.26+311206.3 &  0.00109 & 31.20176 & 0.3622166  & 0.132& 6.281& 58283.4619944& 16.350& 15.844&...  & 0.233& 0.226&  EW   \\
ZTFJ000000.30+233400.5 &  0.00125 & 23.56682 & 0.2698738  & 0.193& 6.302& 58437.2686640& 17.890& 16.944&...  & 0.373& 0.352&  EW   \\
ZTFJ000000.30+711634.1 &  0.00125 & 71.27616 & 0.2685154  & 0.160& 5.236& 58657.4235171& 19.144& 17.875&...  & 0.173& 0.154&  EW   \\
ZTFJ000000.39+605148.8 &  0.00163 & 60.86358 & 0.5591434  & 0.424& 6.322& 58338.4702019& 19.965& 19.002&...  & 0.797& 0.841&  EW   \\
ZTFJ000000.51+583238.7 &  0.00215 & 58.54409 & 2.9797713  & 0.143& 9.013& 58663.4378649& 16.586& 15.368&...  & 0.082& 0.068&  BYDra\\
ZTFJ000001.00+612832.8 &  0.00420 & 61.47580 & 0.3643510  & 0.285& 6.400& 58385.2702714& 20.640& 19.627&...  & 0.970& 0.681&  EW   \\
ZTFJ000001.37+561504.9 &  0.00571 & 56.25137 & 0.2557214  & 0.312& 6.256& 58471.1392038& 19.416& 18.559&...  & 0.674& 0.623&  EW   \\
ZTFJ000001.74+614940.9 &  0.00729 & 61.82803 & 1.0158276  & 0.370& 6.366& 58349.3156217& 19.094& 18.013&...  & 0.289& 0.228&  EW   \\
ZTFJ000001.75+594739.4 &  0.00732 & 59.79428 & 112.0533126& 0.390& 6.950& 58320.7237689& 18.676& 17.393&...  & 0.062& 0.055&  SR   \\
ZTFJ000001.92+554800.7 &  0.00804 & 55.80020 & 112.3366840& 0.209& 5.763& 58389.1259060& 12.664& 0.0   &...  & 0.114& 0.0  &  SR   \\
ZTFJ000002.02+540550.8 &  0.00844 & 54.09746 & 7.6299744  & 0.677& 3.277& 58472.1363770& 18.176& 17.187&...  & 0.187& 0.151&  RSCVN\\
ZTFJ000002.20+480720.8 &  0.00918 & 48.12246 & 0.3810135  & 0.227& 3.701& 58282.4549819& 17.196& 16.037&...  & 0.107& 0.093&  BYDra\\
ZTFJ000002.21+385226.4 &  0.00921 & 38.87401 & 0.4103992  & 0.168& 6.512& 58274.4798066& 15.265& 14.853&...  & 0.209& 0.211&  EW   \\
ZTFJ000002.27+331411.5 &  0.00947 & 33.23655 & 0.1161226  & 0.069& 3.714& 58367.3685172& 18.781& 17.726&...  & 0.238& 0.197&  RSCVN\\
ZTFJ000002.28+592423.4 &  0.00954 & 59.40651 & 0.3286918  & 0.063& 5.511& 58314.4511431& 20.860& 19.268&...  & 0.280& 0.255&  EW   \\
ZTFJ000002.61+560944.8 &  0.01091 & 56.16245 & 0.2780012  & 0.095& 5.197& 58644.4715685& 16.017& 15.664&...  & 0.074& 0.074&  RSCVN\\
ZTFJ000002.69+640614.8 &  0.01123 & 64.10413 & 0.4084968  & 0.250& 6.617& 58307.4094573& 21.241& 19.953&...  & 0.471& 0.440&  EW   \\
ZTFJ000002.99+611736.0 &  0.01246 & 61.29334 & 0.5534820  & 0.269& 6.289& 58431.2049383& 17.362& 16.649&...  & 0.241& 0.228&  EW   \\
ZTFJ000003.05+410045.6 &  0.01271 & 41.01267 & 0.2701826  & 0.236& 6.311& 58296.3618793& 16.796& 16.055&...  & 0.365& 0.354&  EW   \\
ZTFJ000003.10+722838.4 &  0.01292 & 72.47736 & 0.2000857  & 0.014& 5.677& 58364.3618372& 17.146& 16.473&...  & 0.069& 0.054&  RSCVN\\
ZTFJ000003.13--022548.2 &  0.01306 & -2.43006 & 0.4936916  & 0.431& 3.906& 58456.1669642& 17.510& 17.347&...  & 1.301& 0.902&  RR   \\
ZTFJ000003.17+582138.3 &  0.01325 & 58.36065 & 331.9141625& 0.334& 6.273& 58400.7947631& 13.180& 0.0   &...  & 0.073& 0.0  &  SR   \\
ZTFJ000003.23+543605.4 &  0.01347 & 54.60151 & 0.7863971  & 0.105& 8.826& 58389.2748730& 17.639& 16.527&...  & 0.166& 0.112&  BYDra\\
ZTFJ000003.24+692214.2 &  0.01353 & 69.37062 & 0.3728018  & 0.169& 6.048& 58282.4524423& 15.855& 14.678&...  & 0.398& 0.359&  EW   \\
ZTFJ000003.33+550558.3 &  0.01390 & 55.09955 & 0.2513022  & 0.377& 6.044& 58646.4653817& 20.761& 19.701&...  & 0.765& 0.773&  EW   \\
ZTFJ000003.40+623837.5 &  0.01419 & 62.64377 & 0.2627338  & 0.288& 6.245& 58362.3199849& 21.027& 19.922&...  & 0.573& 0.605&  EW   \\
ZTFJ000003.74+393925.9 &  0.01559 & 39.65722 & 0.2394832  & 0.050& 7.755& 58325.4868272& 17.943& 16.731&...  & 0.192& 0.158&  EW   \\
ZTFJ000003.74+623637.4 &  0.01560 & 62.61040 & 0.3755206  & 0.344& 6.383& 58343.3639966& 20.429& 19.378&...  & 0.653& 0.652&  EW   \\
ZTFJ000003.76+532917.1 &  0.01568 & 53.48811 & 0.5921959  & 0.408& 7.638& 58352.3259739& 16.446& 15.747&...  & 0.164& 0.159&  BYDra\\
ZTFJ000003.88+553759.9 &  0.01619 & 55.63331 & 0.2652778  & 0.341& 6.266& 58348.4098300& 19.632& 18.711&...  & 0.747& 0.661&  EW   \\
ZTFJ000003.88+673554.8 &  0.01620 & 67.59858 & 7.8057274  & 0.096& 6.155& 58386.3788529& 20.419& 18.262&...  & 0.498& 0.426&  EW   \\
ZTFJ000003.92+614456.1 &  0.01636 & 61.74894 & 0.3473332  & 0.272& 6.092& 58257.4187066& 20.679& 19.451&...  & 0.521& 0.511&  EW   \\
ZTFJ000003.96+182425.2 &  0.01654 & 18.40700 & 0.4851359  & 0.456& 3.909& 58295.4504510& 15.268& 15.096&...  & 1.126& 0.879&  RR   \\
ZTFJ000004.14+570616.8 &  0.01725 & 57.10468 & 246.2392553& 0.319& 3.591& 58347.9920985& 15.589& 13.508&...  & 1.600& 1.459&  SR   \\
ZTFJ000004.34+554359.8 &  0.01809 & 55.73328 & 0.3470882  & 0.272& 6.349& 58274.4650397& 19.871& 19.026&...  & 0.602& 0.610&  EW   \\
ZTFJ000004.34+522031.5 &  0.01809 & 52.34209 & 4.0833915  & 0.189& 7.602& 58320.3673808& 17.703& 16.745&...  & 0.087& 0.095&  BYDra\\
ZTFJ000004.34+510505.0 &  0.01810 & 51.08474 & 0.3332412  & 0.034& 4.458& 58635.4032937& 17.188& 16.681&...  & 0.192& 0.191&  EW   \\
ZTFJ000004.41+581846.9 &  0.01840 & 58.31303 & 0.4418413  & 0.126& 3.603& 58278.4683106& 18.842& 17.921&...  & 0.110& 0.088&  BYDra\\
ZTFJ000004.49+514615.4 &  0.01871 & 51.77095 & 0.0918632  & 0.187& 4.174& 58428.2433174& 17.392& 17.104&...  & 0.161& 0.115&  DSCT \\
ZTFJ000004.56+453022.7 &  0.01902 & 45.50633 & 6.6168334  & 0.084& 6.243& 58490.2445564& 16.378& 15.673&...  & 0.047& 0.052&  RSCVN\\
ZTFJ000004.98+524230.9 &  0.02076 & 52.70860 & 7.1420745  & 0.359& 6.024& 58471.1790394& 17.357& 15.933&...  & 0.080& 0.072&  BYDra\\
ZTFJ000005.00+555613.9 &  0.02087 & 55.93720 & 0.8834472  & 0.278& 6.004& 58332.3958118& 18.238& 16.738&...  & 0.141& 0.131&  EW   \\
ZTFJ000005.13+503833.5 &  0.02138 & 50.64264 & 6.8118535  & 0.046& 4.986& 58316.3336707& 16.984& 16.128&...  & 0.125& 0.072&  BYDra\\
ZTFJ000005.21+650537.8 &  0.02174 & 65.09385 & 0.1251307  & 0.066& 8.199& 58353.4356974& 14.071& 13.139&...  & 0.107& 0.070&  DSCT \\
ZTFJ000005.34+543148.7 &  0.02228 & 54.53020 & 0.0602120  & 0.101& 4.100& 58305.4711181& 14.514& 14.305&...  & 0.080& 0.055&  DSCT \\
ZTFJ000005.74+573228.7 &  0.02393 & 57.54131 & 3.5412731  & 0.215& 5.945& 58304.4028037& 16.206& 15.416&...  & 0.060& 0.059&  BYDra\\
ZTFJ000005.94+555900.1 &  0.02479 & 55.98338 & 0.4009322  & 0.071& 6.156& 58354.4293181& 17.257& 16.647&...  & 0.167& 0.175&  EW   \\
...                    &  ...     &  ...     &  ...       &  ... &  ... & ...         &  ...  &  ...  &...  &  ... &  ... &  ... \\   
...                    &  ...     &  ...     &  ...       &  ... &  ... & ...         &  ...  &  ...  &...  &  ... &  ... &  ... \\   
...                    &  ...     &  ...     &  ...       &  ... &  ... & ...         &  ...  &  ...  &...  &  ... &  ... &  ... \\   
\hline
\end{tabular}
\tablenotetext{a}{The entire table is available in the online journal;
  50 lines are shown here for guidance regarding its form and
  content.}
\end{center}
\end{table*}

\section{The periodic variables catalog}

\begin{figure}[h!]
\centering
\includegraphics[angle=0,width=88mm]{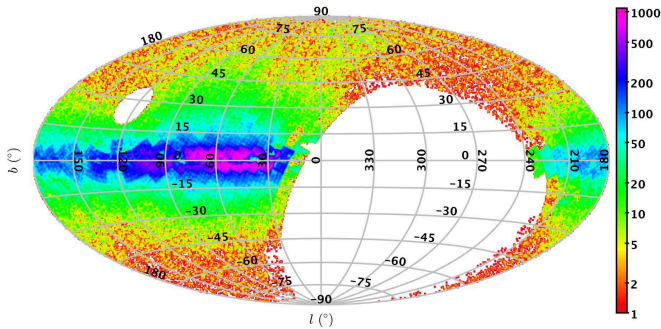}
\caption{\label{map} Distribution map (in Galactic coordinates) of the
  781,602 periodic variables classified in ZTF DR2.}
\end{figure}

The periodic variables catalog\footnote{The full catalog and LCs can
  be accessed at \url{http://variables.cn:88/ztf/} or
  \url{doi.org/10.5281/zenodo.3886372}.} of ZTF DR2, containing 781,602
periodic variables, is included in Table \ref{t2}. It contains the
source ID, position (J2000 R.A. and Dec.), period, mean magnitudes
($\langle g \rangle$, $\langle r \rangle$), number of detections, LC
parameters, and type of variable star. The LC parameters include the
amplitude, amplitude ratio $R_{21}$, phase difference $\phi_{21}$,
Heliocentric Julian Date (HJD) of the minimum $T_0$, $R^2$ of the LC
fit, two-band minimum differences for eclipsing binaries, and the
period's FAP. Candidates without a good classification are collected
in a suspected variables catalog (Table \ref{t6}). The single-exposure
photometry catalog is available online, including the objects'
SourceID, R.A. (J2000), Dec. (J2000), HJD, gmag, rmag, e\_gmag, and
e\_rmag. Figure \ref{map} shows the density distribution of these
periodic variables in Galactic coordinates. This density distribution
is similar to the distribution of the number of detections. 74\% of
the variables are located in the northern Galactic plane
($|b|<10^\circ$), where variables have thus far not been recorded
systematically.

\subsection{New variables}

We cross-matched our variables catalog with several other large
variables catalogs to check the accuracy of our classification and of
the derived periods, as well as the completeness of our catalog. The
reference catalogs include the ASAS-SN Catalog of Variable Stars in
the Northern Hemisphere, the ASAS-SN Catalog of Variable Stars with a
reclassification of 412,000 known variables, the ATLAS catalog of
variables, the {\sl WISE} catalog of periodic variable stars, the
Catalina catalog of periodic variable stars, the {\sl Gaia} catalog of
RR Lyrae and Cepheids, the OGLE variables catalog, and other variables
collected in the General Catalogue of Variable Stars (GCVS). We
assembled these catalogs and then used an angular radius of $2''$ to
match the nearest known variables to our ZTF variables. Almost all
objects were detected within $1''$, which means that the recent
variable surveys all have good positional accuracy.  In addition, a
$1''$ matching radius can significantly reduce sample confusion, which
is a potentially serious problem in dense fields.  621,702 of the
781,602 variables in the ZTF database are newly detected
variables. The total number and the number of newly detected variables
of each type are listed in Table \ref{T3}. The new detection rate of
these variables is high, ranging from 10\% for RRab Lyrae to more than
90\% for $\delta$ Scuti and BY Dra variables. This high new detection
rate is not only the result of the database's high photometric
accuracy ($\sim$0.02 mag accuracy down to $r = 18$ mag) and the faint
limiting magnitude of the ZTF database (about 1.5 mag deeper than
ATLAS in the corresponding $r$ band) but also of the improvements made
in our method of variable-star identification and classification. The
relatively low new detection rate of RRab Lyrae is explained easily,
because they were already well-covered by the Catalina survey and {\sl
  Gaia DR2}. However, both the purity and the period accuracy of the
RRab Lyrae are significantly improved in our ZTF catalog.

\subsection{Classification Accuracy}

The classification accuracy pertaining to each type of variable was
tested by comparison with the four variable-star catalogs: see Table
\ref{T4}. The coincidence rates range from 95\% to 99\%, with any
disagreement usually found at the faint end for each catalog. In
addition, different sampling strategies may also result in somewhat
inconsistent classifications. We note that a few (sub)types have
relatively low coincidence rates of around 80\%. This means that the
standard of classification for these variables is not
well-established. Among the four catalogs we compared our results
with, the ASAS-SN catalog has the best classification accuracy
overall.

In our catalog, RRab Lyrae have the best classification
standard. Their purity is around 99\%. The main contaminants are
unsolved RRc Lyrae and EW-type eclipsing binaries. RRc Lyrae exhibit a
purity of 91\%. Their more symmetric LCs can easily mask them as
EW-type eclipsing binaries. The purity of the eclipsing binaries is
not as high as that of the RRab Lyrae. This is because of the
difficulty in unequivocal classifications among the three binary
subtypes. For lower LC amplitudes the classification becomes rather
difficult. In this paper, we classify all eclipsing binaries into EW
and EA types. Contamination by non-binaries is only 1\%. This means
that our separation of binaries from pulsators is efficient. EW-type
eclipsing binaries have a purity of 96\%; their main contaminants are
EA-type eclipsing binaries. EA-type eclipsing binaries have a
coincidence rate of 84\%.

The coincidence rate of classical Cepheids is around 92\%, based on a
comparison of $\sim$200 known Cepheids. The percentage of disagreement
for Type II Cepheids is around 10\%. LPV purity is around 98\%. Here,
the main misclassification occurs between SRs and Miras, even though
all reference catalogs adopted a similar boundary, $\rm{Amp.}=2.0$
mag. These different classifications can be explained by the different
limiting magnitudes pertaining to the reference catalogs: a shallow
limiting magnitude will lead to underestimated Mira amplitudes. For
example, the ASAS-SN catalog has a limiting magnitude of 17, so that
for Miras fainter than 15 mag, the amplitude determined is likely less
than 2 mag. These Miras will hence be classified as SRs and, as a
result, about 46\% of matched Miras are mistaken for SRs in the
ASAS-SN catalog. In turn, 99.2\% of ASAS-SN Miras are correctly
classified in our catalog. Toward the faint magnitude limit of ZTF,
this type of misclassification also occurs. In our catalog, 1694 SRs
with $m_{g,r}>19$ mag and ${\rm Amp}_r>1.0$ (${\rm Amp}_g>1.2$) mag
run the risk of misclassification. 

The short-period $\delta$ Scuti variables have a purity of
88\%. However, $\delta$ Scuti stars are not well classified in
previously published catalogs. In our catalog, the purity of HADS is
as high as those of RR Lyrae and eclipsing binaries, while the LADS
purity is relatively lower. As regards rotational stars, the purity of
BY Dra (93.7\%) is high since they were strictly selected based on
their {\sl Gaia} parallaxes. However, without a luminosity criterion,
it is not easy to separate RS CVn and eclipsing binaries with
poor-quality photometry. To obtain a cleaner sample of eclipsing
binaries, many eclipsing binaries with relatively poor-quality LCs
have been classified as RS CVn, which leads to a systematically lower
purity level of RS CVn (75.2\%). Given that we have double checked the
LCs of all variables where we noticed disagreement with other
catalogs, our classification is more reliable for most of these
objects. This is reasonable, since ZTF is deeper, has better-quality
photometry, and has been obtained in two passbands. However, toward
the faint end of the ZTF, the ZTF sample purity will also decrease.

\begin{table}[h!]
\vspace{-0.0in}
\begin{center}
\caption{\label{T3}All and newly detected ZTF variables.}
\vspace{0.15in}
\begin{tabular}{lcr}
\hline
\hline                                                
Type         &   Total    &   New (fraction)     \\
\hline                                 
Cep-I          &      1262   &     565   (44.8\%)  \\   
Cep-II         &       358   &     154   (43.0\%)  \\   
RRab           &    32,518   &    3034   ( 9.3\%)  \\   
RRc            &    13,875   &    2178   (15.7\%)  \\   
$\delta$ Scuti &    16,709   &  15,396   (92.1\%)  \\   
EW             &   36,9707   & 306,375   (82.9\%)  \\   
EA             &    49,943   &  40,201   (80.5\%)  \\   
Mira           &    11,879   &   4,997   (42.1\%)  \\   
SR             &   119,261   &  97,737   (82.0\%)  \\   
RS CVn         &    81,393   &  70,957   (87.2\%)  \\   
BY Dra         &    84,697   &  80,108   (94.6\%)  \\   
Total number   &   781,602   & 621,702   (79.5\%)  \\  
\hline
\end{tabular}
\end{center}
\end{table}

\begin{table*}[ht!]
\vspace{-0.0in}
\begin{center}
\caption{\label{T4} Variable purity comparison between ZTF and other catalogs.}
\vspace{0.15in}
\begin{tabular}{lccccc}
\hline
\hline
Type            & ATLAS           & ASAS-SN        & Catalina        & WISE           &  Total sample  \\
\hline
RRab            &  99.8\%         & 98.6\%         &  98.8\%         & 96.2\%         &   99.0\%        \\
RRc             &  89.0\%         & 90.4\%         &  93.9\%         &                &   91.2\%        \\
EW              &  95.1\% (99.8\%)\tablenotemark{b} & 95.1\% (97.1\%)  &  97.1\% (97.7\%)& 96.3\% (98.4\%)   &   95.6\% (98.7\%) \\
EA              &  82.0\% (99.8\%) & 93.6\% (96.0\%) &  70.5\% (98.3\%) & 95.6\% (98.5\%) &   84.3\% (99.2\%) \\
Cep-I           &  98.5\%         & 84.1\%         &                 & 96.1\%         &   91.7\%        \\
Cep-II          &  96.9\%         & 82.8\%         &  94.5\%         & 88.1\%         &   90.1\%        \\
SR              &  97.7\%         & 98.0\%         &                 &                &   97.9\%        \\
Mira            &  88.5\%         & 53.3\%         &                 &                &   69.3\%        \\
$\delta$ Scuti  &  83.5\%         & 94.6\%         &  85.5\%         &                &   87.6\%        \\        
RS CVn          &                 & 75.2\%         &                 &                &   75.2\%        \\
BY Dra          &                 & 93.7\%         &                 &                &   93.7\%        \\
\hline
\end{tabular}
\tablenotetext{b}{The purities in brackets only account for contamination by non-binary variables.}
\end{center}
\end{table*}

\subsection{Period Accuracy}

\begin{figure}[htbp!]
\centering
\vspace{-0.0in}
\includegraphics[angle=0,width=80mm]{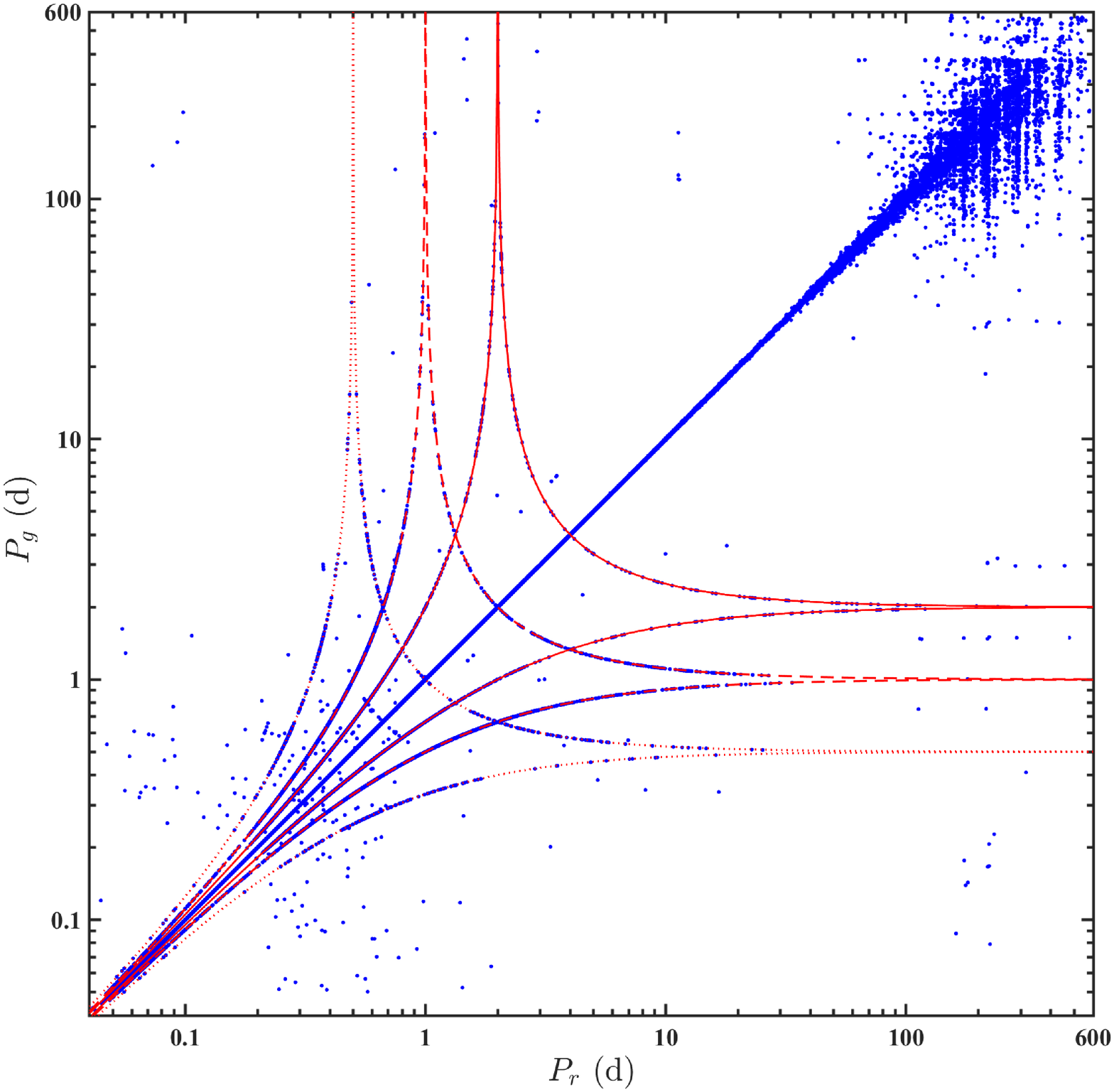}
\vspace{-0.0in}
\caption{\label{pp} Comparison of periods determined from the $g$- and
  $r$-band LCs. The red lines are the predicted relations between real
  and aliased periods.}
\end{figure}

ZTF DR 2 covers a timespan of 470 days, so the periods in our catalog
are most accurate for $P\lesssim 100$ days. This limitation causes
problems for the periods of SRs and Miras. Period problems pertaining
to short-period variables are usually caused by the limited sampling
cadence, e.g., a daily cadence is a major problem for ground-based
time-domain surveys. Except for the {\sl WISE} catalog, the other
three previously published catalogs and the ZTF catalog are all
affected by this problem. This period problem is not easily eliminated
by increasing the sampling using the same cadence, but it can be
reduced based on multi-passband information and information about the
physical properties of the variables. Figure \ref{pp} shows a
comparison of the periods determined from the $g$ and $r$ LCs. One of
the two periods is the real period ($P_{\rm r}$), while the other is
the aliased period ($P_{\rm a}$). Both periods are found to mainly
follow three relations: $1/P_{\rm a}=|1/P_c\pm1/P_{\rm r}|, P_c=0.5,
1, 2$ (corresponding to the red dotted, dashed, and solid lines in
Figure \ref{pp}, respectively). We can justify which period is more
reliable, e.g., the bright variables are usually LPVs ($M_{W_{gr}}<-2$
mag), or variables with $\phi_{21}<4$ are likely associated with RRab
rather than RRc periods. However, for eclipsing binaries and $\delta$
Scuti stars, there is no significant feature to validate the real
period. For example, assuming that $\delta$ Scuti stars have real and
aliased periods of 0.08 and 0.074 days, respectively, it is hard to
justify which period is real, even with access to their luminosities
and $\phi_{21}$ phases. Therefore, we visually compared the folded LCs
pertaining to the two periods individually for those variables. This
process can determine the correct period for most variables if the LC
quality is good.

Our comparison of the four catalogs shows that our period accuracy is
as high as 99\%, except for LPVs and $\delta$ Scuti stars. The
inconsistency rates for RRab, RRc, EW-type eclipsing binaries, and for
classical Cepheids, range from 0\% to 2\% (Table \ref{t5}). This is a
common accuracy for large variable-star catalogs. The inconsistency
rate of $\delta$ Scuti stars is a little higher, since their periods
are much shorter than the sampling cadence. Complementary
short-cadence observations can resolve this aliased period
problem. The full- versus half-period problem occurred for a few
EA-type eclipsing binaries with insignificant secondary eclipses. If
we do not consider the full- versus half-period problem, the period
accuracy of EA-type eclipsing binaries is as high as those for other
short-period variables. We have visually checked the LCs of EA-type
eclipsing binaries to ensure that they all show the full periods. As
to LPVs, around 11\% of Miras and 32\% of SRs have period differences
larger than 20\% by comparison with the ASAS-SN and Catalina
catalogs. The precision of Mira periods is limited by the
observational timespan. SRs usually contains several short- and
long-term periods, so the low inconsistency derived here is not
surprising. The accuracy of the LPV periods will be refined with the
availability of future DRs.

\begin{table*}[htbp!]
\vspace{-0.0in}
\begin{center}
\caption{\label{t5} Period comparison between ZTF and other catalogs.}
\vspace{0.15in}
\begin{tabular}{lccccc}
\hline
\hline
Type            & ATLAS           & ASAS-SN        & Catalina        & {\sl WISE}           &  Total sample  \\
\hline
RRab            &  99.7\%        & 99.7\%         &  98.0\%          & 99.6\%          &   99.6\%        \\
RRc             &  98.9\%        & 98.4\%         &  97.8\%          &                &   98.3\%        \\
EW              &  99.8\%        & 98.6\%         &  95.3\%          & 99.9\%        &    98.7\% \\
EA              &  92.9\% (98.7\%)\tablenotemark{c} & 93.8\% (99.4\%)  &  95.1\% (98.1\%)   & 93.1\% (99.7\%)   &   93.2\% (98.9\%) \\
Cep-I           &  100\%         & 97.9\%         &                 & 100\%           &   99.2\%        \\
SR              &               & 68.3\%         &  93.5\%          &                &   68.4\%        \\
Mira            &               & 88.2\%         &  94.6\%        &                &   88.7\%      \\
$\delta$ Scuti   &  98.8\% & 94.8\%  &  91.5\%   &                &   96.6\% \\  
\hline
\end{tabular}
\tablenotetext{c}{The values in brackets do not take into account
  the full- versus half-period problem.}
\end{center}
\end{table*}

\subsection{Completeness}

We estimate the completeness of our catalog by comparison with
previously published catalogs based on known variables in the ZTF
magnitude range deemed reliable ($r>12.5$ mag), as well as its spatial
coverage ($\rm{Dec.}>-13^\circ$). The completeness for the full sample
is 71.1\%. Since sampling of the ZTF is not homogeneous across the
entire northern sky, the completeness varies with spatial position
(see Figure \ref{map}). For regions at high declinations
($\rm{Dec.}>40^\circ$), the completeness can be as high as 76\%. Some
24\% of our variables are not classified because of insufficient
sampling. The ZTF DR2 contains $\sim 150$ detections for each object,
a smaller number than the equivalent numbers of detections in the
ASAS-SN and Catalina catalogs. As the declination decreases, the
completeness gradually decreases to 50\%. For different magnitude
ranges, the completeness is always around 71.1\%, which means that the
completeness is not affected by the objects' apparent magnitudes or
photometric uncertainties. With future DRs, ZTF will be able to detect
and classify more than 1 million periodic variables in the northern
sky, assuming current completeness levels. {\sl Gaia} may be able to
help classify 3 million periodic variables across the full sky,
considering their enhanced density in the Galactic bulge and the
Magellanic Clouds in the southern sky.

Except for spatial sampling, time sampling issues may also cause
incompleteness. Since our catalog only covers a timespan of 470 days,
its completeness for LPVs is not as high as that for short-period
variables. We compare the LPVs in the ASAS-SN catalog with our sample
to estimate their completeness, since ASAS-SN covers both a longer
timespan and a larger number of LPVs. Globally, 42\% of LPVs across
the entire sky were rediscovered in our catalog. This fraction is some
30\% lower than that for our full sample of variables.

\subsection{Cepheids and RR Lyrae}

In addition to our comparisons with the four variables catalogs, we
also evaluated the periods and completeness levels of the Cepheids and
RR Lyrae in our catalog by comparison with the OGLE and {\sl Gaia}
catalogs, both of which have similar depths as ZTF. OGLE IV recently
released their variables catalog of Cepheids and RR Lyrae in the
Galactic disk \citep{2019AcA....69..305S,
  2019AcA....69..321S}. Although the overlap region between OGLE and
ZTF is limited, a comparison is still feasible. {\sl Gaia} DR2
contains about 140,000 RR Lyrae across the full sky, as well as 500
classical Cepheids in the Milky Way.

As for the classical Cepheids, we compared our catalog with the 2682
high-probability Cepheids collected by the OGLE team
\citep{2013AcA....63..379P,2018AcA....68..315U}. The number of objects
in common is 691, and the completeness of our ZTF Cepheids is 63.3\%
given that 1092 of the full sample of Cepheids are located in the
ZTF's detection region. This relatively low completeness level
compared with the global completeness of our variables catalog
suggests that the completeness of variable types with intermediate
periods is lower than that of short-period variables. Only two of the
691 confirmed objects have inconsistent periods. Having double checked
these periods, we conclude that our periods are more reliable.

OGLE IV contains 10,000 RR Lyrae in the Galactic disk, and two of
their fields overlap with the field covered by the ZTF catalog. We
found that 1327 of the 1854 (71.6\%) OGLE RR Lyrae in these two fields
have been rediscovered in our catalog; 99.5\% of these 1327 RR Lyrae
have consistent periods. Both the period accuracy and completeness are
in agreement with the equivalent values pertaining to the full
catalog. For RR Lyrae contained in {\sl Gaia} DR2, 39,301 are located
in the ZTF's detection region. Some 80.1\% (31483) of {\sl Gaia} RR
Lyrae have been rediscovered in our catalog. This completeness level
is 9\% higher than the catalog's global completeness of 71.1\%, which
is mainly driven by the incompleteness levels characteristic of {\sl
  Gaia} RR Lyrae. About 400 of the {\sl Gaia} RR Lyrae are
contaminated by EW-type eclipsing binaries and $\delta$ Scuti stars.

\section{Discussion}

In this section, we discuss for each type of variable their LCs, PLRs,
and their applications as potential distance tracers.

\subsection{Cepheids}

\begin{figure*}[ht!]
\centering
\hspace{0.0in}
\includegraphics[angle=0,width=160mm]{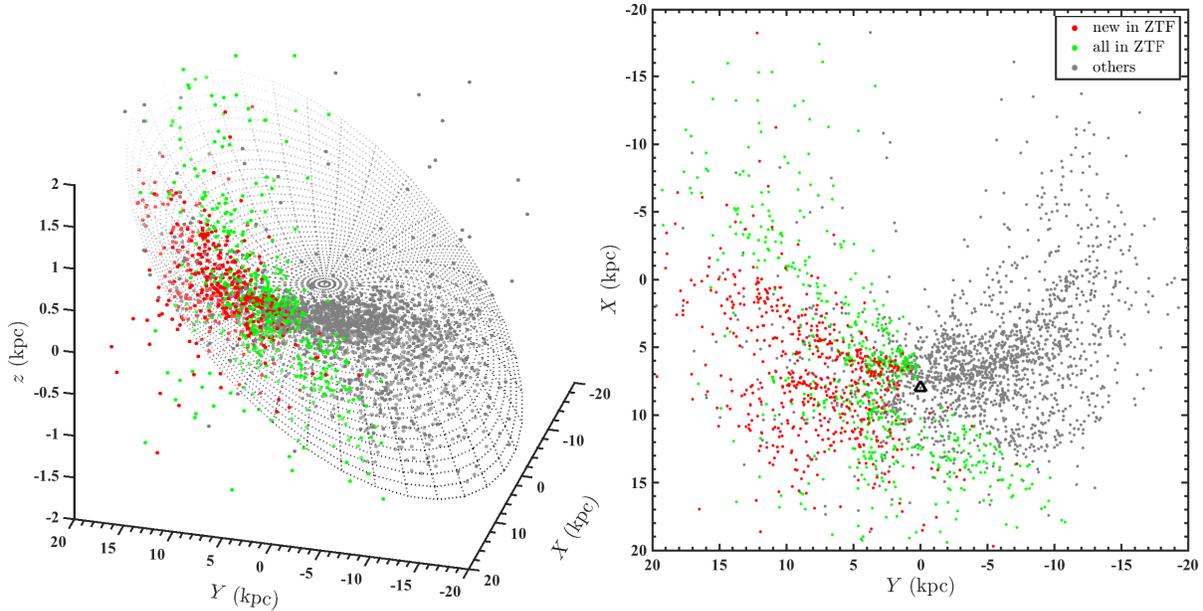}
\vspace{-0.0in}
\caption{\label{cepdisk} Milky Way disk in 3D (left) based on
    $\sim$3300 Cepheids and their projection onto the $(X,Y)$ plane
    (right). Red: new Cepheids in the ZTF catalog; green: other
    Cepheids in ZTF catalog; gray: Cepheids from OGLE, {\sl WISE},
    ATLAS, ASAS-SN, and other catalogs. The grid in the left-hand
    panel represents the warp model of \cite{2019NatAs...3..320C}. The
    triangle in the right-hand panel denotes the position of our
    Sun.}
\end{figure*}

Classical Cepheids are the most important primary distance
tracers. They are widely used for estimating distances to galaxies in
the Local Volume and measuring the near-field Hubble constant
\citep{2019ApJ...876...85R}. Because of their young age, classical
Cepheids are also the best stellar tracer of the Galactic
disk. Compared with extragalactic Cepheids, Cepheids in our Milky Way
are rare. Before 2018, only around 1000 Milky Way Cepheids had been
detected \citep{2005AcA....55..275P, 2008yCat.2285....0B}. These
Cepheids were distributed within 3 kpc of the Sun and showed few
features of our Milky Way's structure. In 2018, several times the
number of previously known classical Cepheids were found. The {\sl
  WISE} catalog of variables listed about 1000 Cepheids that were
homogeneously distributed across the Galactic plane, except for the
Galactic Center direction. OGLE detected another one thousand Cepheids
in the southern disk. ALTAS, ASAS-SN, and {\sl Gaia} also detected
several hundred Cepheids.

This rapid increase in the number of Cepheids motivated studies of the
Milky Way's disk. Samples of 1339--2500 carefully selected disk
Cepheids revealed the first intuitive 3D map of our Galaxy's stellar
disk \citep{2019NatAs...3..320C, 2019Sci...365..478S}. The outer disk
was found to be warped, and the warp is now known to be precessing
\citep{2019NatAs...3..320C}. Recently, 640 new Cepheids in the
southern disk affected by heavy extinction were discovered by the VVV
survey \citep{2019ApJ...883...58D}. In our ZTF catalog, we found an
additional 565 new classical Cepheids, thus further enlarging the
sample in the northern warp and around the edge of the disk. Figure
\ref{cepdisk} shows the distribution of 3300 well-determined disk
classical Cepheids ( satisfying the {\sl Gaia} parallax and LC
criteria), which agree with the warp model of
\citet{2019NatAs...3..320C} to within 1--2$\sigma$. This new sample of
Cepheids is sufficient to trace the detailed morphology of the disk
beyond Galactocentric distances of 15 kpc. In the $(X,Y)$ plane, the
Cepheid distribution is not homogeneous, but it shows bubble and
filament features. A significant bubble is found at a distance of
about 6 kpc from the Sun in the Galactic Anticenter direction. With a
detailed study of these features, we could potentially infer the
dynamical evolution history of the Galactic disk.

In the future, new Cepheids are expected to be found in the Galactic
Center direction \citep{2011Natur.477..188M, 2015ApJ...812L..29D}. In
addition, low-amplitude or long-period Cepheids will be classified
once more accurate parallaxes become available. Our larger sample of
Cepheids will also be helpful in determining the zero points and
slopes of their PLRs \citep{2019arXiv191004694B}, particularly of
infrared PLRs \citep{2017MNRAS.464.1119C, 2018ApJ...852...78W}. In
addition to Cepheids in the Milky Way, we also found 21 Cepheids (one
new Cepheid) associated with M31 \citep{2018AJ....156..130K}, one
Cepheid in M33 \citep{2006MNRAS.371.1405H}, and one in IC 1613
\citep{2001AcA....51..221U}. ZTF DR2 contains only about 50 detections
in the M31 direction. With increased sampling in future DRs, several
hundred Cepheids will likely be found in M31.

\begin{figure}
\centering
\hspace{0.0in}
\includegraphics[angle=0,width=88mm]{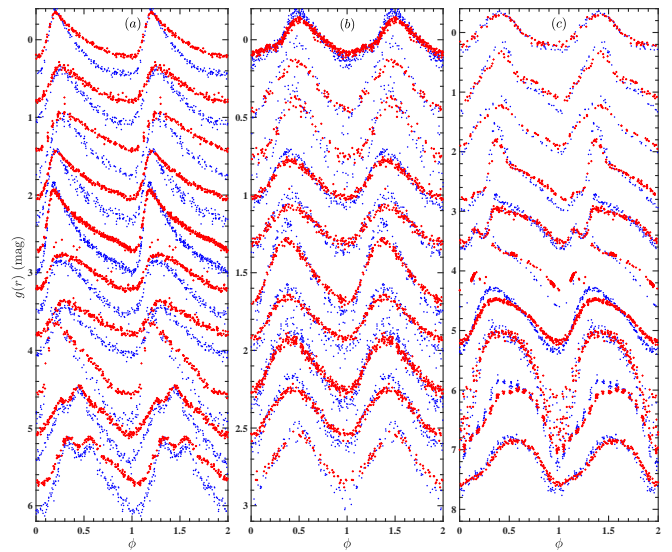}
\vspace{-0.0in}
\caption{\label{ceplc} Example LCs for (a) fundamental-mode classical
  Cepheids, (b) first-overtone classical Cepheids, and (c) Type II
  Cepheids. The blue dots and red plus signs are LCs in the $g$ and
  $r$ bands, respectively. From top to bottom, phase differences
  $\phi_{21}$ increase. The mean magnitudes are fixed at $0.6i+0.2$,
  $0.3i$, and $0.8i$ ($i=0, 1, ..., 9$) in the $g$ band and $0.6i$,
  $0.3i$, and $0.8i$ in the $r$ band, for fundamental-mode classical
  Cepheids, first-overtone classical Cepheids, and Type II Cepheids,
  respectively.}
\end{figure}

Cepheid LCs in $g$ and $r$ are shown in Figure \ref{ceplc}. From their
LCs, the majority of classical Cepheids can be divided into
fundamental-mode (Figure \ref{ceplc}a) and first-overtone Cepheids
(Figure \ref{ceplc}b). First-overtone Cepheids have lower amplitudes
and shorter periods than the corresponding fundamental-mode
Cepheids. At a given period, first-overtone Cepheids have slightly
brighter luminosities. This magnitude difference is around 0.5 mag in
the LMC \citep{2016ApJ...832..176I}, Small Magellanic Cloud
\citep[SMC][]{2017MNRAS.472..808R}, and M31
\citep{2018AJ....156..130K}, while the situation in the Milky Way is
less clear \citep{2019A&A...625A..14R}. With an increasing number of
first-overtone Cepheids and better parallaxes, the PLRs of Milky Way
Cepheids will be better constrained.

Compared with classical Cepheids, the LC shapes of Type II Cepheids
are more highly variable (Figure \ref{ceplc}c). Type II Cepheids are
older, fainter, and follow PLRs characterized by a somewhat larger
scatter \citep{2017A&A...604A..29G, 2017A&A...605A.100B}. Type II
Cepheids are mostly located away from the disk and associated with the
Galactic bulge and halo. As distance indicators, Type II Cepheids can
be used to study the structure of the bulge
\citep{2018A&A...619A..51B} and measure the distances to globular
clusters and dwarf galaxies. We found 154 new Type II Cepheids,
which represents a significant increase and will be useful to anchor
their PLRs and explore old Galactic structures.

\subsection{RR Lyrae}

\begin{figure}
\centering
\hspace{0.0in}
\includegraphics[angle=0,width=88mm]{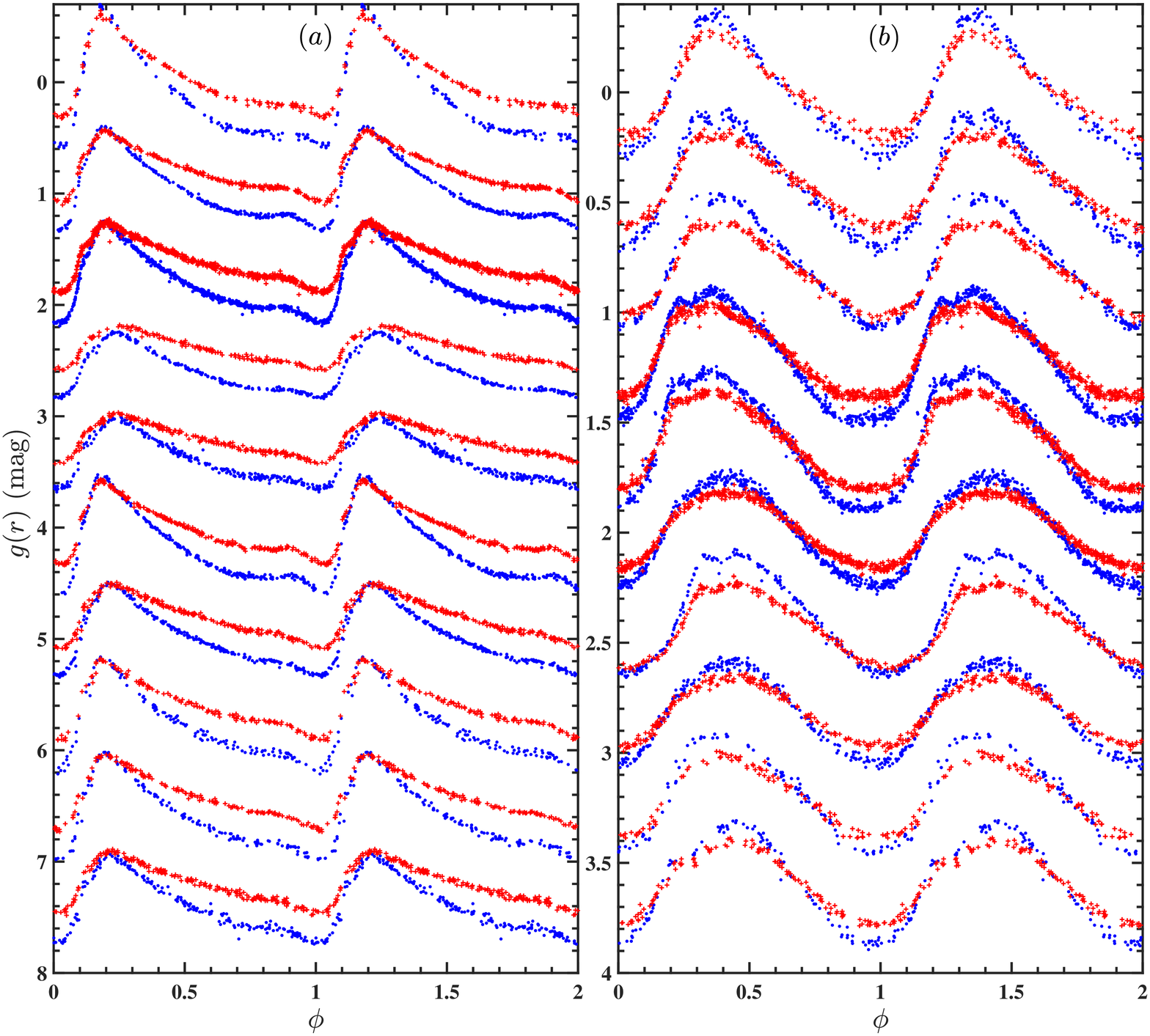}
\vspace{-0.0in}
\caption{\label{rrlc} Example LCs for (a) RRab and (b) RRc Lyrae. The
  blue dots and red plus signs are LCs in the $g$ and $r$ bands,
  respectively. From top to bottom, phase differences $\phi_{21}$
  decrease.}
\end{figure}

RR Lyrae are the most numerous distance indicators in old
environments. Their PLRs or PL--metallicity relations can be as
accurate as those for the classical Cepheids \citep[][and references
  therein]{2009Ap&SS.320..261C}. In the Milky Way halo, RR Lyrae are
used to trace dwarf galaxies and distant substructures
\citep{2013ApJ...765..154D, 2017ApJ...844L...4S, 2018ApJ...855...43M,
  2019MNRAS.490.2183M, 2019MNRAS.487.2685E}. RR Lyrae can be used to
reveal structures out to distances of some 120 kpc. Better
identification of RR Lyrae in the Dark Energy Survey can help to reach
distances greater than 200 kpc \citep{2019AJ....158...16S}. RR Lyrae
at moderate distances are usually used to determine accurate distances
to globular clusters \citep{2018AJ....155..137B, 2019ApJ...870..115B,
  2019MNRAS.487.3140P}. With a more complete sample, RR Lyrae are
perhaps the best tracers to trace the Milky Way halo's shape
\citep{2018MNRAS.474.2142I, 2019MNRAS.482.3868I}. In the Galactic
inner regions, featuring numerous RR Lyrae, the shape of the bulge can
also be constrained \citep{2015ApJ...811..113P,
  2016A&A...591A.145G}. In the ZTF catalog, we detected more than
5,000 new RR Lyrae out to distances of 125 kpc. We detected several RR
Lyrae associated with the Draco dwarf galaxy at a distance of around
76 kpc \citep{2004AJ....127..861B}. Unlike previous RR Lyrae samples,
the new sample may reveal unknown structures at low Galactic latitudes
($|b|<20^\circ$).

Studies of the physical properties of RR Lyrae will also benefit from
large and complete samples. In recent years, many metal-rich RR Lyrae
have been found \citep{2017ApJ...835..187C, 2018AJ....155...45S},
which challenge the traditional theory explaining these metal-poor and
old variables. Both the {\sl WISE} and ZTF catalogs focus on the
Galactic disk; RR Lyrae in the disk are, with high probability,
metal-rich. With LAMOST or SDSS-SEGUE spectra, the physical properties
of a large sample of RR Lyrae can be investigated. In Figure \ref{pa},
the distribution of RR Lyrae in the period--amplitude diagram is more
scattered than dichotomous, which suggests that the well-known
Oosterhoff dichotomy problem of RR Lyrae may just be due to a lack of
intermediate-metallicity RR Lyrae \citep{2019ApJ...882..169F}.

Similarly to classical Cepheids, most RR Lyrae can be divided into
fundamental-mode (RRab) and first-overtone pulsators (RRc). Example
LCs of RRab and RRc Lyrae are shown in Figure \ref{rrlc}. The
amplitudes of RRab Lyrae are larger, decreasing as the LCs become
symmetric (from top to bottom in the left-hand panel of Figure
\ref{rrlc}). This trend is the result of the period--amplitude
relation and is only significant for RRab LCs. The PLRs or
PL--metallicity relations of Milky Way RRab Lyrae can be studied with
{\sl Gaia} DR2 parallaxes \citep{2018MNRAS.481.1195M,
  2019AJ....158..105L, 2019MNRAS.tmp.2412N}. The accuracy of these
PLRs is mainly limited by parallax uncertainties. RRc Lyrae also
follow PLRs, which are about 0.5 mag brighter than for RRab Lyrae at a
given period. In Figure \ref{pl}, RRc Lyrae seem to exhibit more
scatter in their luminosities than RRab Lyrae. If we can better
constrain the luminosities of RRc Lyrae, they may be potentially
viable distance tracers. A number of RRab and RRc Lyrae exhibit
multiple periods, and about one-third of the RRab Lyrae show period
modulation. These special RR Lyrae will be analyzed with enhanced data
from future DRs.

\subsection{Eclipsing binaries}

\begin{figure}[h!]
\centering
\hspace{0.0in}
\includegraphics[angle=0,width=85mm]{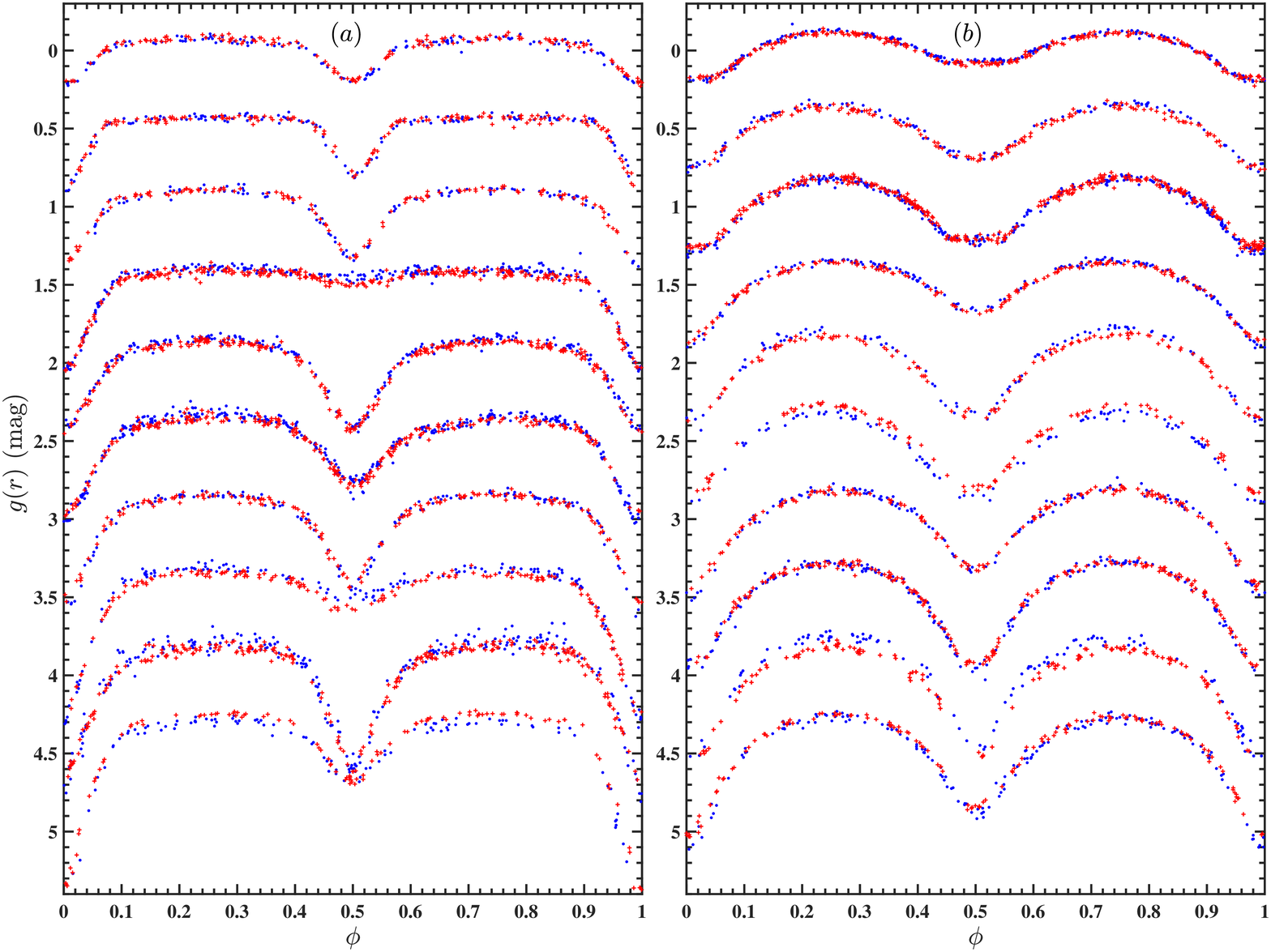}
\vspace{-0.0in}
\caption{\label{eclc} Example LCs for (a) EA- and (b)
  EW-type eclipsing binaries. The blue dots and red plus signs are LCs
  in the $g$ and $r$ bands, respectively. From top to bottom, the
  variation amplitudes increase.}
\end{figure}

The three types of eclipsing binaries are distinguished by their
evolutionary stage. From EA- to EW-type eclipsing binaries, the period
distribution gradually shifts to shorter periods. Statistically,
EW-type eclipsing binaries tend to be older, with a peak in their
  age distribution around 4 Gyr. EA-type eclipsing binaries are
young, with a peak age around 1 Gyr. The basic evolution of
eclipsing binaries has recently become clear. Nuclear evolution and
angular momentum loss are the main mechanisms
\citep{1988MNRAS.231..341H, 2001ApJ...552..664N, 2005ApJ...629.1055Y,
  2006AcA....56..199S, 2020MNRAS.492.2731J}. Both mechanisms run in
parallel. Driven by angular momentum loss, the period decreases and
both components become closer. When one component evolves to or away
from the terminal main sequence, its radius increases. Until one
component fills its Roche lobe, an EA-type eclipsing binary evolves to
become an EB-type eclipsing binary. Then, with material and energy
transferring from the primary to the secondary component, combined
with possible mass reversal, the secondary component also fills its
Roche lobe. The resulting system is an EW-type eclipsing binary. In
Figure \ref{cla1}, the significant cut-off period of EW-type eclipsing
binaries around 0.19 days is driven by the nuclear evolution
timescale. The progenitor of the primary component of this 0.19-day
EW-type eclipsing binary is a low-mass star with a nuclear evolution
timescale very close to the age of the Galaxy.

Since their binary fraction is higher, eclipsing binaries account for
a large proportion of variables in magnitude-limited samples. The
fraction of EW-type eclipsing binaries was estimated at around 0.1\%
in the field \citep{2006MNRAS.368.1319R}. It could be as high as 0.4\%
in a $\sim4$ Gyr environment \citep{2016AJ....152..129C}. This
fraction does not correct for the low-inclination problem and only
counts EW-type eclipsing binaries seen at higher inclinations
($i>40$--$60^\circ$). In our catalog, the fraction of EW-type
eclipsing binaries identified is 0.07\%, which is a lower limit. If we
consider the number of missing EW-type eclipsing binaries owing to
insufficient sampling (see Section 5.4), that fraction can be as high
as 0.1\%.

Detached eclipsing binaries evolve on timescales as long as or even
longer than those of EW-type eclipsing binaries. However, depending on
the spatial separation and radius ratio of both components, we can
only observe eclipses of detached eclipsing binaries with
$i\sim=90^\circ$. Consequently, the fraction of EA-type eclipsing
binaries is lower than that of EW types. EB-type eclipsing binaries
have the shortest evolution timescale and account for the lowest
fraction among the three types of eclipsing binaries.

Figure \ref{eclc} shows LCs of eclipsing binaries. Their shapes and
amplitudes in both bands are almost the same and differ from those of
pulsating stars. In the left-hand panel, we can identify when the
eclipses start in the LCs of EA-type eclipsing binaries. In the
right-hand panel, the eclipse onset in EW(EB)-type eclipsing binary
LCs is hard to identify. In the period range of 0.25--0.56 days,
EW-type eclipsing binaries follow tight infrared PLRs
\citep{2018ApJ...859..140C} and can be potential distance
indicators. The viability of these PLRs has been validated by
eclipsing binaries from the ASAS-SN catalog
\citep{2019arXiv191109685J}.

In Figure \ref{pl}, about 90,000 EW-type eclipsing binaries with
$<20$\% {\sl Gaia} parallax uncertainties are found to also follow a
PLR. In theory, the PLR of EW-type eclipsing binaries is driven by
both Roche lobe theory and the mass--luminosity relation. Based on
this PLR and a sample of several 100,000 EW-type eclipsing binaries, a
better study of the detailed structure of the Milky Way will be
possible. Except for distance determination, EW-type eclipsing
binaries can also provide age information based on their periods. In
addition, kinematic properties derived from {\sl Gaia} data can
provide constraints on the binaries' ages
\citep{2019arXiv190906375H}. We expect that the eight-dimensional
parameter space pertaining to EW-type eclipsing binaries can be
established and used to constrain the Milky Way's evolution. Without
strict constraints on the secondary components, the PLR of EB-type
eclipsing binaries exhibits more significant scatter. Less evolved,
EA-type eclipsing binaries are the best objects to obtain absolute
parameters from the orbital equations. By analyzing both their LCs and
radial velocity curves, late-type EA-type eclipsing binaries can yield
distances with 1\% accuracy \citep{2019Natur.567..200P}.

\begin{figure}[h!]
\centering
\hspace{0.0in}
\includegraphics[angle=0,width=88mm]{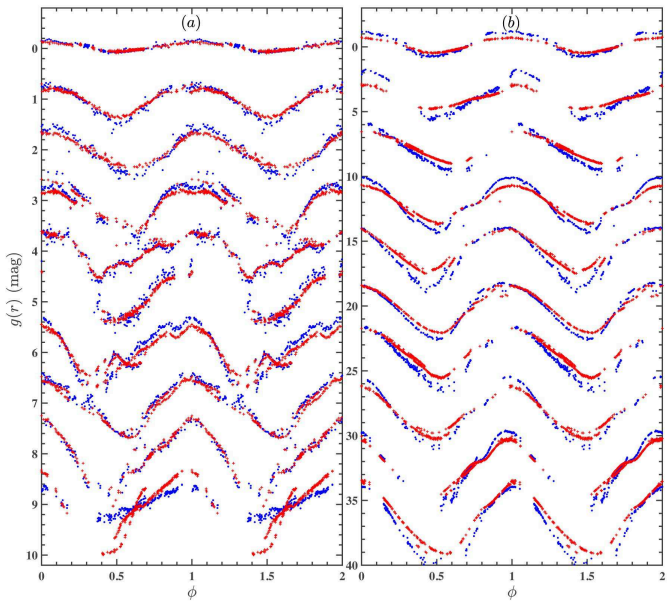}
\vspace{-0.0in}
\caption{\label{lpvlc} Example LCs for (a) SRs and (b) Miras. The blue
  dots and red plus signs are LCs in the $g$ and $r$ bands,
  respectively. From top to bottom, the variation amplitudes
  increase. The mean magnitudes are fixed at $1.0i$ and $4.0i$ ($i=0,
  1, ..., 9$) in the $g$ and $r$ bands, for SRs and Miras,
  respectively.}
\end{figure}

\subsection{SRs and Miras}

SRs and Miras are most likely red giants, red supergiants, or AGB
stars exhibiting long-period luminosity variations. They are important
distance tracers in the context of the cosmological distance scale
since they are brighter than classical Cepheids in infrared
bands. However, LPVs follow multi-sequence PLRs, and each PLR sequence
exhibits larger scatter \citep{2019A&A...631A..24L} than the classical
Cepheid PLRs. To become a reliable distance indicator, LPV properties
must be studied in more detail, particularly in the Milky Way and
M31. The PLRs of red supergiants belonging to the LMC, SMC, M31, and
M33 have been well-studied, and their 1$\sigma$ accuracy is 10--15\%
in infrared bands
\citep{2011ApJ...727...53Y,2012ApJ...754...35Y,2019ApJS..241...35R}. However,
in the Milky Way, the number of red supergiants with reliable
distances is too small to establish reliable PLRs
\citep{2019MNRAS.487.4832C}.

Oxygen-rich Miras also follow tight PLRs. In infrared bands, the
1$\sigma$ scatter about their PLRs is around 10\%
\citep{2017AJ....154..149Y, 2018AJ....156..112Y}. Compared with red
supergiants, Miras are 0--4 mag fainter but several times more
numerous. All of these LPVs are potential tracers to measure the
Hubble parameter \citep{2019arXiv190810883H}. In the Milky Way, LPVs
are young and intermediate-age structure tracers, complementary to
Cepheids and RR Lyrae. Similarly to eclipsing binaries, the ages of
each LPV subtype can be inferred from their periods
\citep{2018A&A...618A..58M}. As such, LPVs are also tracers that can
constrain the evolution history of our Galaxy.

\begin{figure}[h!]
\centering
\hspace{0.0in}
\includegraphics[angle=0,width=88mm]{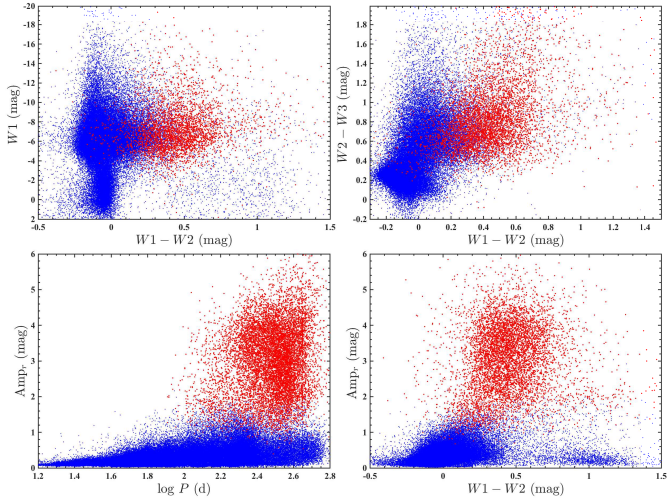}
\vspace{-0.0in}
\caption{\label{lpvd} LPVs in the ZTF catalog. (a) Color--absolute
  magnitude diagram. (b) Color--color diagram. (c) Period--amplitude
  diagram. (d) Color--amplitude diagram. SRs and Miras are shown as
  red and blue dots, respectively.}
\end{figure}

LCs of SRs and Miras are shown in the left- and right-hand panels of
Figure \ref{lpvlc}, respectively. In the ZTF, several Miras are found
with $r$-band amplitudes exceeding 6 mag. Since the total timespan of
the ZTF DR2 is 470 days, many Miras only have LCs available for at
most one period. We detected about 130,000 LPVs, with the majority
belonging to our Galaxy or to nearby dwarf galaxies. To better
understand the physics of these LPVs, we also show them in
color--magnitude and color--color diagrams (Figure
\ref{lpvd}a,b). {\sl WISE} $W_1$, $W_1-W_2$, and $W_2-W_3$ magnitudes
with good photometric signal-to-noise ratios (${\rm SNR}_{W1}>5, {\rm
  SNR}_{W2}>5, {\rm SNR}_{W3}>10$) and quality were adopted to
indicate LPV magnitudes and colors. This choice reduces the effects of
the significant extinction in the Galactic plane. By comparison with
massive stars in the LMC in color--absolute magnitude space
\citep{2018A&A...616A.175Y}, we find that SRs (blue dots in Figure
\ref{lpvd}) are composed of red giants, red supergiants, oxygen- and
carbon-rich AGB stars, as well as extremely carbon-rich AGB
stars. Miras (red dots) contain a few oxygen-rich, a few extremely
carbon-rich, and many carbon-rich AGB stars.

AGB populations are characterized by an absolute magnitude of about
$-8$ mag in the mid-infrared, and the bright end of the red supergiant
population is around $-12$ mag. Therefore, data points with magnitudes
brighter than $-12$ mag in the left-hand panel ($W_1-W_2<0$) or
brighter than $-9$ mag in the right-hand panel ($W_1-W_2>0$) are stars
that may have larger parallax uncertainties. In Figure \ref{lpvd}c,
the amplitudes grow monotonously as the periods increase. In
particular, some sequences are found in the SR distribution, referred
to as the A, B, and C$'$ (third, second, and first overtone) sequences
\citep{2017ApJ...847..139T}. Miras and some SRs form sequence C, which
is the location of high-amplitude fundamental pulsators. In Figure
\ref{lpvd}d, SRs and Miras are separated by a boundary, with Miras
being high-amplitude and red. The detailed subtype classification of
these LPVs is not further discussed in this paper as it requires
access to better periods and parallaxes.

\begin{figure}[h!]
\centering
\hspace{0.0in}
\includegraphics[angle=0,width=88mm]{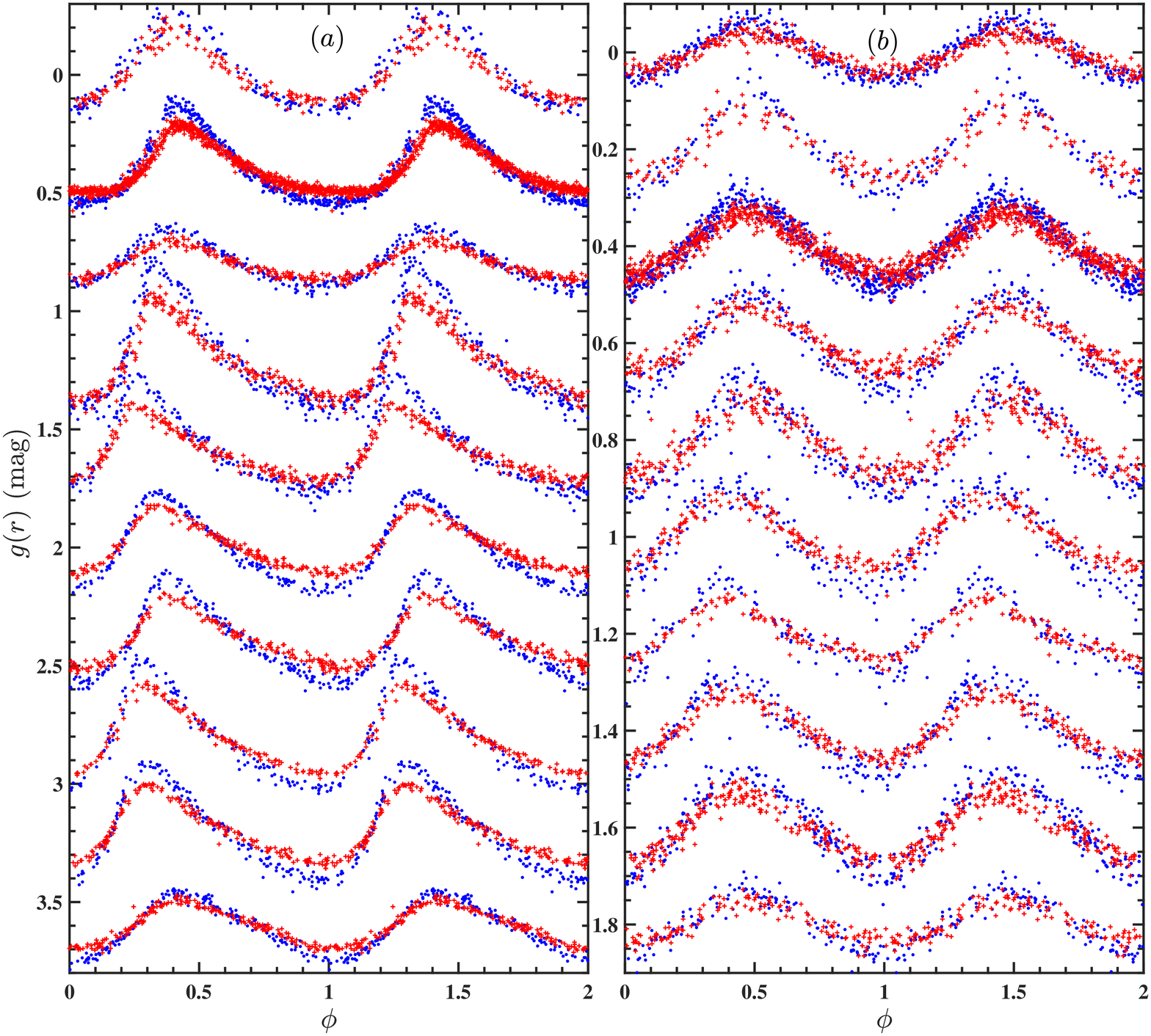}
\vspace{-0.0in}
\caption{\label{dslc} Example LCs for (a) HADs and (b) LADs. The blue
  dots and red plus signs are LCs in the $g$ and $r$ bands,
  respectively. From top to bottom, phase differences $\phi_{21}$
  increase. The mean magnitudes are fixed at $0.4i$ and $0.2i$ ($i=0,
  2, ..., 9$) in the $g$ and $r$ bands, for HADs and LADs,
  respectively.}
\end{figure}

\subsection{$\delta$ Scuti stars}

Among the variables in the ZTF, $\delta$ Scuti stars have the shortest
periods. $\delta$ Scuti stars are main-sequence pulsators located in
the instability strip, so they follow a PLR similar to RR Lyrae and
Cepheids. Different from RR Lyrae and Cepheids, a fraction of $\delta$
Scuti stars is thought to exhibit multiple periods. The PLRs of
$\delta$ Scuti stars are not well studied, even for HADs, because of
their fainter magnitudes and limited sampling in the past. In the LMC,
the magnitude range of $\delta$ Scuti stars ranges from 20 to 22 mag
in the $R$ band \citep{2010AJ....140..328G}, which is close to the
limiting magnitude of many time-domain surveys. PLRs of $\delta$ Scuti
stars are therefore best studied in the Milky
Way. \citet{2019MNRAS.486.4348Z} tried to establish a $V$-band PLR
based on 1100 $\delta$ Scuti stars in the {\sl Kepler} field with good
{\sl Gaia} DR2 parallaxes; however, their PLR is not as tight as the
equivalent PLRs pertaining to other tracers. Recently,
  \citet{2020MNRAS.493.4186J} derived multi-band PLRs for fundamental
  and overtone pulsators, based on $\sim$ 8400 $\delta$ Scuti stars
  from ASAS-SN, and found a 1$\sigma$ scatter around 10\%.

Our ZTF sample is several times larger and more homogeneously
distributed across the northern sky, which paves the way to improve
the $\delta$ Scuti PLRs. In Figure \ref{pl}, we show that the
Wesenheit PLR of $\delta$ Scuti stars is tight, with few outliers. A
simple estimation shows that the 1$\sigma$ scatter of the HADs is
around 10\%. This means that the $\delta$ Scuti PLR is tighter in
infrared bands. This is similar to the situation for other variables
that follow PLRs. A forthcoming paper will investigate the multi-band
PLRs of a large sample of $\delta$ Scuti stars. As distance
indicators, $\delta$ Scuti stars cover a wide age range (according to
their masses). Old $\delta$ Scuti (SX Phoenicis) stars can be used to
measure the distances to globular clusters and nearby dwarf galaxies
\citep{2011AJ....142..110M}. Young or intermediate-age $\delta$ Scuti
stars are found in open clusters \citep{2015AJ....150..161W}, and they
are potential tracers for studies of the Milky Way disk's structure.

In Figure \ref{dslc}, HADs exhibit tooth-saw-shaped LCs similar to RR
Lyrae and Cepheids (left-hand panel), while LADs show various types of
LCs. Based on their LCs (Figures \ref{pa} and \ref{pr21}) alone, it is
difficult to separate fundamental-mode and first-overtone $\delta$
Scuti stars. However, similarly to RR Lyrae and classical Cepheids,
first-overtone $\delta$ Scuti stars are generally brighter than
fundamental-mode $\delta$ Scuti stars. The boundary between HADs and
LADs is not clear. In the period--luminosity diagram (Figure
\ref{pl}), LADs also follow a PLR and are a little brighter than
HADs. All these features imply that HADs and LADs are first-overtone
and fundamental-mode $\delta$ Scuti stars, respectively, while their
boundary is rather obscure. More properties of $\delta$ Scuti stars
will be revealed by high-order frequency analysis with increased
sampling in the next ZTF DR.

\begin{figure}[ht!]
\centering
\hspace{0.0in}
\includegraphics[angle=0,width=88mm]{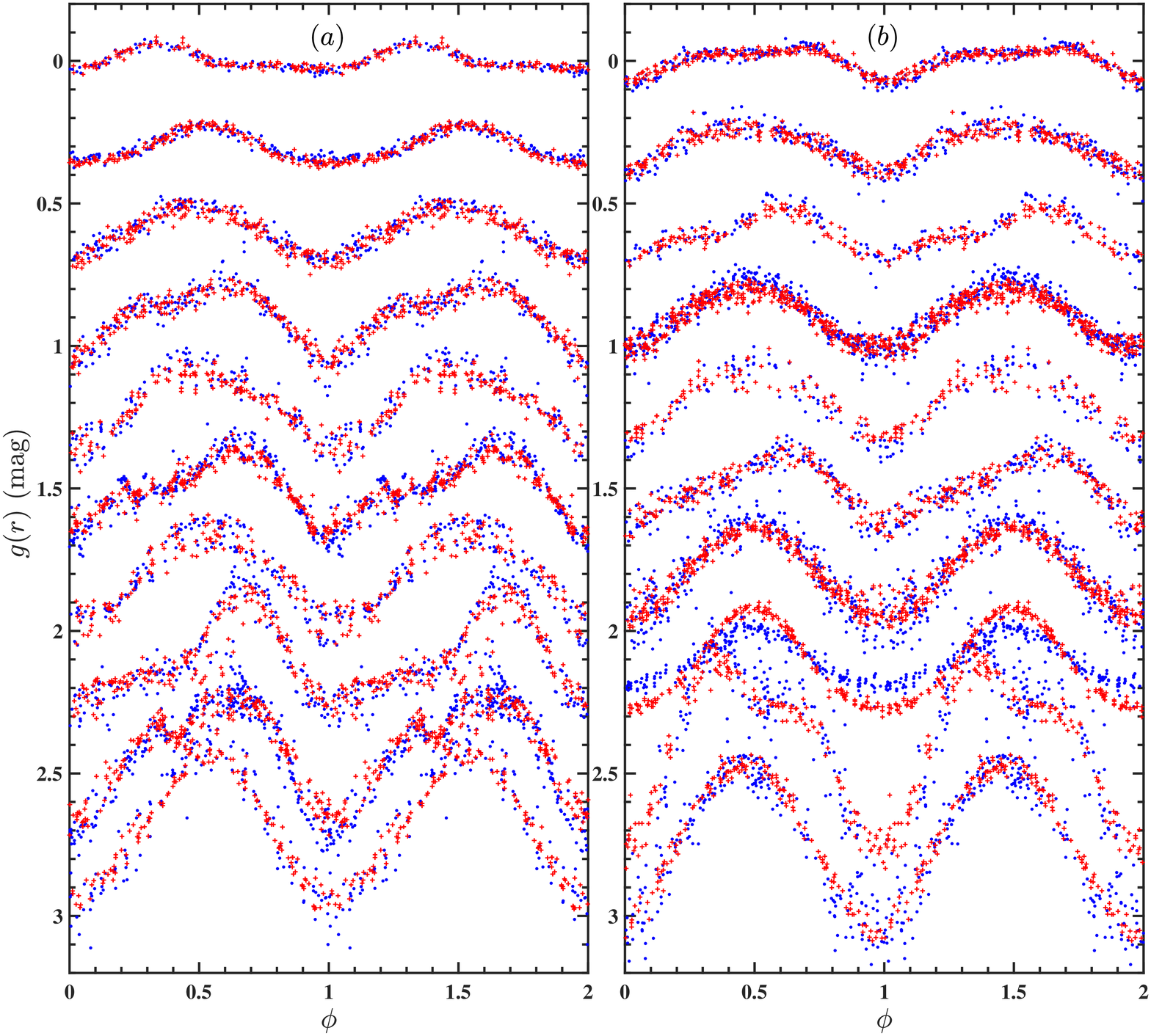}
\vspace{-0.0in}
\caption{\label{rotlc} Example LCs for (a) RS CVn and (b) BY Dra
  stars. The blue dots and red plus signs are LCs in the $g$ and $r$
  bands, respectively. From top to bottom, the variation amplitudes
  increase. The mean magnitudes are fixed at $0.3i$ ($i=0, 1, ..., 9$)
  in the $g, r$ bands for both RS CVn and BY Dra stars.}
\end{figure}

\subsection{RS CVn and BY Dra}

Both RS CVn and BY Dra exhibit periodic but non-characteristic LCs
(Figure \ref{rotlc}). Their amplitudes or luminosities can be used to
classify them. The LCs of RS CVn and BY Dra in the two different bands
are similar. However, the overall shapes of their LCs are no longer
symmetric about phase $\phi=0.5$ or $\phi=1.5$ (in most cases). A
number of these variables have been detected in the ASAS-SN Catalog of
the Southern Hemisphere \citep{2020MNRAS.491...13J}, where the authors
treat them as rotational stars. \citet{2019ApJ...879..114I} also
  studied 12,660 spotted variable stars along the line of sight to and
  inside the Galactic bulge, based on OGLE data. Most of their
  variables are giant stars. In this paper, we divided these
variables based on their different luminosities. RS CVn are F--G-type
eclipsing binaries with strong chromospheric activity. Their spectra
show Ca {\sc ii} H and K emission \citep{1976ASSL...60..287H}. The LCs
of RS CVn are a combination of eclipses and spot modulation. RS CVn
have thus far only been studied based on small sample sizes
\citep{2014MNRAS.442.2620Z}. Therefore, our catalog of $\sim 80,000$
RS CVn is the best sample to investigate these objects' properties
statistically. A new discovery already resulting from this enlarged
sample is that RS CVn have a wider period distribution than previously
inferred \citep{1980AcA....30..387H, 1993ApJS...86..599D,
  1997MNRAS.291..658D}. In Figure \ref{cla2}, for all periods,
eclipsing binaries have the highest probability of having RS CVn
features. Nevertheless, the peak of the RS CVn $\log P$ distribution
occurs around a few days, which is much longer than that for eclipsing
binaries (0.3--0.4 days).

BY Dra stars are also variables featuring chromospheric activity, but
they are of K--M type and fainter than RS CVn. Several papers have
studied BY Dra and some RS CVn stars based on samples of a few hundred
objects \citep{1993A&AS..100..173S, 2008MNRAS.389.1722E,
  2011AJ....141..166H}. However, criteria to separate the two types of
variables are not well-established. In our sample, with the help of
{\sl Gaia} parallaxes, BY Dra stars are predominantly distributed in a
narrow magnitude range, $2.8<M_{W_{gr}}<4.6$ mag (Figure
\ref{pl}). The period range of BY Dra stars is narrower than that of
RS CVn variables. As for the overall distributions of periods and
amplitudes, both types of variables exhibit similar features. A better
study of RS CVn and BY Dra variables will be achieved with a
combination of LCs and spectra.

\begin{table*}[ht!]
\tiny
\vspace{-0.0in}
\begin{center}
\caption{\label{t6}ZTF Suspected Variables Catalog.}
\vspace{0.15in}
\begin{tabular}{lccccccccccc}
\hline
\hline
   ID               & R.A. (J2000)  & Decl. (J2000)  &  $\langle g \rangle$ & $\langle r \rangle$ & Period$_g$& Period$_r$ &Amp$_g$&Amp$_r$ & Num$_g$ &Num$_r$ & $\log$ (FAP)$_g$\\
                      &$^\circ$&$^\circ$& mag &   mag  &  days     & days       &  mag   &   mag  &        &        &                 \\
\hline
ZTFJ000000.06+583327.2&0.00026&58.55756&12.974&12.324  & 1.00157   &  0.05415   &  0.059 &  0.057 &  118   &  137   &  $-$5.865\\                                           
ZTFJ000000.09+632559.3&0.00041&63.43315&15.137&14.451  & 339.5882  &  344.55948 &  0.075 &  0.038 &  124   &  146   &  $-$6.685\\                                           
ZTFJ000000.14+552150.6&0.00060&55.36408&17.081&15.919  & 0.49772   &  0.09674   &  0.049 &  0.024 &  121   &  143   &  $-$5.067\\                                           
ZTFJ000000.15+540322.4&0.00066&54.05624&18.144&16.776  & 0.49768   &  0.10488   &  0.087 &  0.023 &  122   &  144   &  $-$5.566\\                                           
ZTFJ000000.16+621059.7&0.00069&62.18326&16.454&15.585  & 0.4979    &  0.08946   &  0.046 &  0.018 &  125   &  146   &  $-$5.494\\                                           
ZTFJ000000.17+592103.3&0.00073&59.35093&17.342&16.391  & 0.99392   &  0.13364   &  0.055 &  0.026 &  124   &  147   &  $-$4.545\\                                           
ZTFJ000000.25+605628.0&0.00108&60.94112&17.745&15.993  & 0.99385   &  0.44865   &  0.127 &  0.019 &  125   &  146   &  $-$3.811\\                                           
ZTFJ000000.27+665503.0&0.00116&66.91750&16.711&15.505  & 251.12632 &  0.05878   &  0.069 &  0.021 &  169   &  186   &  $-$6.459\\                                           
ZTFJ000000.28+583810.1&0.00119&58.63614&15.121&14.359  & 295.46337 &  0.33318   &  0.068 &  0.059 &  118   &  138   &  $-$5.534\\   

\hline
\end{tabular}
\tablecomments{(This table is available in its entirety in machine-readable form.)}
\end{center}
\end{table*}

\section{Conclusions}
ZTF DR2 contains about 0.7 billion objects with more than 20 exposures
each. As such, ZTF DR2 represents a highly suitable database for the
detection and exploration of new variable stars. In this paper, we
have adopted the Lomb--Scargle periodogram approach to search for
periodic variables. We obtained a sample of 1.4 million candidates
with $\rm{FAP}<0.001$. Following exclusion of candidates with aliased
periods and poor-quality LCs, $\sim 1.0$ million candidates remained
for classification. Classification was done using the DBSCAN method
based on the distributions of the objects' periods, LC parameters, and
luminosities. We classified 781,602 periodic variables into $\delta$
Scuti, EW- and EA-type eclipsing binaries, RRab and RRc Lyrae,
classical Cepheids, Type II Cepheids, SRs, and Miras. After
cross-matching with known variables, 621,702 of the 781,602 objects
turned out to be newly detected variables. This high new detection
rate highlights both the high quality of the ZTF photometry and the
improvement of our classification method.

Checks of the periods and types contained in our catalog show that
they are better than or as good as the same parameters contained in
previously published catalogs. The sampling completeness is not
homogeneous in the northern sky due to the observation mode adopted,
ranging from 76\% at high declinations to 50\% in low declination
regions ($\rm{Dec.}<0^\circ$). The completeness of LPVs is 30\% lower
than that of the full sample. For each variable type, our catalog
represents a significant increase in sample size. The new variables
will aid studies of the physics of variable stars as such, of the
Milky Way's structure and evolution, and of the cosmological distance
scale, particularly based on a combination of LCs and spectra.

Future ZTF DRs will cover both increased timespans and larger numbers
of exposures. These aspects will help improve the determination of the
LPV periods and increase the completeness of the catalog. LCs
associated with double modes or amplitude (period) modulation will be
re-analyzed when enhanced sampling becomes available. In addition to
the main types of variables discussed here, there are many cataclysmic
variables, low-amplitude pulsating stars, binaries with compact
objects, and stars with exoplanets in the ZTF database. These
variables will be identified to construct a large sample for further
study based on future ZTF DRs.

\acknowledgments{We thank the anonymous referee for comments that
  helped us improve the paper. This publication is based on
  observations obtained with the Samuel Oschin 48-inch Telescope at
  the Palomar Observatory as part of the Zwicky Transient Facility
  project. ZTF is supported by the National Science Foundation under
  grant AST-1440341 and a collaboration including Caltech, IPAC, the
  Weizmann Institute for Science, the Oskar Klein Center at Stockholm
  University, the University of Maryland, the University of
  Washington, Deutsches Elektronen-Synchrotron and Humboldt
  University, Los Alamos National Laboratories, the TANGO Consortium
  of Taiwan, the University of Wisconsin at Milwaukee, and Lawrence
  Berkeley National Laboratories. Operations are conducted by COO,
  IPAC, and UW. We are grateful for research support from the National
  Natural Science Foundation of China through grants 11903045,
  U1631102, and 11633005. This work was supported by the National Key
  Research and Development Program of China through grant
  2017YFA0402702. We also thank Dr. Changqing Luo and Dr. Fan Yang for providing
  computing time.}

\bibliography{ztf2_vari}{}
\bibliographystyle{aasjournal}

\end{document}